\documentclass[a4paper,11pt]{article}

\usepackage{jheppub} 

\usepackage[T1]{fontenc} 
\usepackage{amssymb,amsfonts,amsmath}
\usepackage{graphicx}
\usepackage{pdfpages}
\usepackage{color} 
\newcommand{\cred}[1]{{\color{red}#1}}

\newcommand{\bea}{\begin{eqnarray}}
\newcommand{\eea}{\end{eqnarray}}

\newcommand{\epi}{\epsilon_i}
\def\gsim{\mathrel{
   \rlap{\raise 0.511ex \hbox{$>$}}{\lower 0.511ex \hbox{$\sim$}}}}
\def\lsim{\mathrel{
   \rlap{\raise 0.511ex \hbox{$<$}}{\lower 0.511ex \hbox{$\sim$}}}}

\title{\boldmath $h \rightarrow \gamma \gamma$ in $U(1)_{R}-$ lepton number 
model with a right-handed neutrino}

\author[a,1]{Sabyasachi Chakraborty,\note{Corresponding author.}}
\author[b]{AseshKrishna Datta,}
\author[a]{Sourov Roy}

\affiliation[a]{Department of Theoretical Physics, Indian Association for
the Cultivation of Science, 2A $\&$ 2B Raja S.C.Mullick Road, Jadavpur,
Kolkata 700 032, INDIA}
\affiliation[b]{Harish-Chandra Research Institute, Chhatnag Road, Jhunsi, 
Allahabad 211019, INDIA}

\emailAdd{tpsc3@iacs.res.in}
\emailAdd{asesh@hri.res.in}
\emailAdd{tpsr@iacs.res.in}

\cred{
\preprint{
\begin{flushright}
HRI-P-14-10-001
\end{flushright}
}}

\abstract{We perform a detailed study of the signal rate of the lightest Higgs boson 
in the diphoton channel ($\mu_{\gamma \gamma}$), recently analyzed by both the 
ATLAS and CMS collaborations at the Large Hadron Collider, in the framework 
of $U(1)_R-$ lepton number model with a right handed neutrino superfield. 
The corresponding neutrino Yukawa coupling, `$f$', plays a very important  
role in the phenomenology of this model. A large value of 
$f\sim\mathcal O(1)$ provides an additional tree level contribution to the 
lightest Higgs boson mass along with a very light (mass $\sim$ a few hundred 
MeV) bino like neutralino and a small tree level mass of one of the active 
neutrinos that is compatible with various experimental results. In the presence 
of this light neutralino, the invisible decay width of the Higgs boson can become
important. We studied this scenario in conjunction with the recent LHC results.
The signal rate $\mu_{\gamma\gamma}$ obtained in this scenario is compatible 
with the recent results from both the ATLAS and the CMS collaborations 
at 1$\sigma$ level. A small value of `$f$', on the other hand, is compatible 
with a sterile neutrino acting as a 7 keV dark matter that can explain the observation 
of a mono-energetic X-ray photon line by the XMM-Newton X-ray observatory. We
also study the impact of $\mu_{\gamma\gamma}$ in this case. 
}

\begin{document}
\maketitle
\flushbottom


\section{Introduction}
Recently two CERN based Large Hadron Collider (LHC) experiments, ATLAS and CMS, 
have confirmed the existence of a neutral boson, widely accepted to be the 
Higgs boson, an elementary scalar boson of nature \cite{ATLAS-higgs, CMS-higgs}, 
with mass around $125$ GeV. Almost all the decay channels have been probed with 
reasonable precision. Out of these, results in the $h\rightarrow \gamma\gamma$ 
channel have attracted a lot of attention in recent times. The reason is two-fold: 
first, this is the discovery mode of the Higgs boson and second, being a 
loop induced process it may potentially carry indirect hints of new physics. The 
results reported so far show some deviations with respect to the Standard Model 
(SM) prediction. For example, the ATLAS collaboration reported $\mu_{\gamma\gamma} 
= 1.17\pm 0.27$ \cite{ATLAS-mugg-1}, where $\mu_{\gamma\gamma}=\frac{\sigma
(pp\rightarrow h\rightarrow \gamma\gamma)}{\sigma (pp\rightarrow h\rightarrow 
\gamma\gamma)^{SM}}$. On the other hand, CMS collaboration reported a 
best-fit signal strength in their main analysis \cite{CMS-mugg-2} where, 
$\mu_{\gamma\gamma} = 1.14^{+ 0.26}_{- 0.23}$. Moreover, a cut-based analysis 
by CMS produced a slightly different value, which is quoted as 
$\mu_{\gamma\gamma}=1.29^{+ 0.29}_{- 0.26}$. This enhancement or suppression in
the $h\rightarrow\gamma\gamma$ channel with respect to the SM provide a natural 
testing ground for physics Beyond the SM (BSM). Detailed studies have already 
been carried out for this particular channel. For example, $h\rightarrow\gamma
\gamma$ is studied in a wide variety of supersymmetric (SUSY) models namely, the 
minimal supersymmetric standard model (MSSM) \cite{Arbey,Bechtle,Winkler,Drees,
Arbey-1,Staub,Low,Carena,Hall,Heinemeyer,Arbey-2,Draper,Chen,Ren,Djouadi,Yuan,
Gori,Su,Moretti,Moretti-1,Amit}, its next-to-minimal version (NMSSM) \cite{King,Gunion,
Belanger-1,Gunion-1,Hugonie,Ellwanger,Cao,Kang,Huitu}, the constrained MSSM (CMSSM) 
\cite{Kadastik,Baer,Aparicio,Ellis,Barger,Cao-J} and also in (B-L)SSM \cite{Elsayed,
Basso,Moretti-3,Khalil}, left-right supersymmetric models \cite{Huitu-1}, and in 
$U(1)^{\prime}$ extension of MSSM \cite{Subha}. 
In \cite{Mohanty}, a triplet-singlet extension of MSSM has been studied 
and $\mu_{\gamma\gamma}$ is computed. 

Motivated by these results we would like to investigate the Higgs to diphoton 
mode in the context of a supersymmetric scenario known as $U(1)_{R}-$ lepton 
number model, which is augmented by a single right-handed neutrino superfield. 
It is rather well known that supersymmetry is one of the very popular frameworks 
that provides a suitable dark matter candidate and can also explain the origin of 
neutrino masses and mixing. However, the non-observation of superpartners so far 
has already put stringent lower bounds on their masses in different SUSY models, 
subject to certain assumptions. In the light of these constraints, $R$-symmetric 
models which generically contain Dirac gauginos in their spectra (as opposed to 
Majorana gauginos in usual SUSY scenarios) are very well motivated. In particular, 
the presence of Dirac gluino in this class of models reduces the squark production 
cross section compared to MSSM thus relaxing the bound on squark masses. Detailed
studies on $R$-symmetric models and Dirac gauginos can be found in the literature
~\cite{Fayet,Polchinski,Hall-1,Hall-2,Jack,Nelson,Fox,Chacko,Antoniadis,Antoniadis-1,
kribs,Choi,Amigo,Benakli,Belanger,Benakli-m,Kumar,Fox-1,Benakli-m2,Choi-1,Carpenter,
Kribs-1,Abel,Benakli-2,Kalinowski,Benakli-1,Gregoire,Katz,Rehermann,Davies,
ItoyamaMaru,Bertuzzo,Davies-1,Argurio,Fok,Argurio-1,Kumar-1,Claudia,Goodsell,Riva,
Chakraborty,Csaki,Dudas,Beauchesne,Bertuzzo:2014bwa,Benakli:2014cia,Roy-Ghosh,
Goodsell:2014dia,Ipek:2014moa,Busbridge:2014sha,Diessner:2014ksa}. Flavor and CP violating 
constraints are also suppressed in these class of models \cite{kribs}. To 
construct Dirac gaugino masses, the gauge sector of the supersymmetric Standard 
Model has to be extended to incorporate chiral superfields in the adjoint 
representations of the SM gauge group. A singlet $\hat S$, an $SU(2)$ triplet 
$\hat T$ and an $SU(3)$ octet $\hat O$, help obtain the Dirac gaugino masses. 

In this paper we consider the minimal extension of a specific $U(1)_{R}$ 
symmetric model \cite{Kumar-1,Claudia} by introducing a right 
handed neutrino superfield \cite{Chakraborty}. In such a scenario the R-charges 
are identified with lepton numbers such that the lepton number of SM fermions 
and their superpartners are negative of the corresponding R-charges. Such an 
identification leaves the lepton number assignments of the SM fermions unchanged 
from the usual ones while the same for the superpartners become non-standard. 
This has an interesting consequence for the sneutrinos which now do not carry 
any lepton number. Hence, although in this model sneutrinos get non-zero vacuum 
expectation value ($vev$) in general, the latter do not get constrained from 
neutrino Majorana masses which require lepton number violation by two units. A 
sneutrino thus can play the role of a down type Higgs boson, a phenomenon which 
has crucial implications \cite{Gregoire,Bertuzzo,Kumar-1,Claudia,Chakraborty} 
for our purpose that we would discuss later in this work. The right handed 
neutrino, on the other hand, not only provides a small tree level Dirac neutrino 
mass but also gives rise to an additional tree level contribution to the Higgs 
boson mass proportional to the neutrino Yukawa coupling \cite{Chakraborty}. When 
the R-symmetry is broken, a small ($\lsim$ 0.05 eV) Majorana mass for one of 
the active neutrinos is generated at the tree level while the right handed sterile 
neutrino can have keV Majorana mass and can be accommodated as a warm dark matter 
candidate \footnote{For a review on other models of keV sterile neutrino dark matter, 
see~ref. \cite{Merle}.}.

A large Yukawa coupling $f\sim \mathcal O(1)$ facilitates having the mass of 
the lightest Higgs boson around $125$ GeV without resorting to radiative 
contributions. Large values of $f$ also result in a very light neutralino with 
mass around a few hundred MeV. Cosmological implications of having such a light 
neutralino is briefly discussed in ref. \cite{Chakraborty} for this model. Some 
general studies regarding light neutralinos can be found in~\cite{Dreiner,Hooper,Lebedev,
Ota,Kim,Dreiner:2009er,Choudhury,Adhikari,Adhikari-1,Adhikari-2}. On the other hand, in the 
regime of small Yukawa coupling $f\sim 10^{-4}$, the Higgs boson mass is devoid 
of any large tree level contribution. Therefore, to obtain the mass of the 
lightest Higgs boson in the right ballpark, radiative corrections have to be 
incorporated, which are required to be large enough. This can be achieved either 
by having large singlet and triplet couplings~\cite{Belanger}, $\lambda_{S}$, 
$\lambda_{T}$ $\sim O(1)$, or by having a large top squark mass. 

In this work, we study the implications of such a scenario with particular 
reference to the diphoton final states arising from the decay of the lightest 
Higgs boson. We study this scenario in conjunction with the recent results of
$\mu_{\gamma \gamma}$ obtained from the latest results of LHC collaborations.
This particular case under consideration has some important implications since we can 
now afford rather light top squarks which potentially affect the resonant production 
rate of the lightest Higgs boson and its decay pattern. Furthermore, presence of 
a very light neutralino opens up new decay modes of the Higgs bosons which in 
turn is subject to the constraints from Higgs invisible branching fractions. Also, 
in general, presence of new particle states and their involved couplings would 
affect the proceedings.

The plan of the work is as follows. In Section 2 we briefly discuss the main 
features of the model. The principal motivation and the artifacts of the 
$U(1)_{R}-$ lepton number model are also discussed with reference to its scalar 
and the electroweak gaugino sector. In section 3, we discuss the scalar sector
of the model in detail. In Section 4 we address the neutralino and 
the chargino sectors. The masses and the couplings in these sectors 
play important roles in the computation of $\mu_{\gamma\gamma}$. A thorough 
analysis of $\mu_{\gamma\gamma}$ requires the knowledge of both production and 
decays of the Higgs boson. In Section 5 issues pertaining to the production of 
Higgs boson in the present scenario is discussed in some detail. Analytical 
expressions of Higgs boson decaying to two photons in our model are also given in 
the same section. Section 6 is dedicated to the computation of the invisible
decay width of the Higgs boson. Here we also discuss the impact of the
findings from the LHC pertaining to the Higgs sector on the scenario under discussion
for two distinct cases: a) when the neutrino Yukawa coupling is large, i.e.,
$\mathcal O(1)$ and b) when it is $\mathcal O(10^{-4})$.
We also provide  $\mu_{\gamma\gamma}$ 
and show its variation with relevant parameters, along with the points representing 
the 7 keV sterile neutrino warm dark matter in this model. 
We conclude in Section 7 with some future outlooks. The Higgs boson couplings 
to neutralino and charginos in this model are relegated to the appendix.

\section{$U(1)_R$-lepton number model with a right handed neutrino}
We consider a minimal extension of an $R$-symmetric model, first discussed  in 
\cite{Kumar-1,Claudia}, by extending the field content with a 
single right handed neutrino superfield \cite{Chakraborty}. Along with the MSSM 
superfields, $\hat H_{u}$, $\hat H_{d}$, $\hat U_{i}^c$, $\hat D_{i}^c$, 
$\hat L_{i}$, $\hat E_{i}^c$, two inert doublet superfields $\hat R_{u}$ and 
$\hat R_{d}$ with opposite hypercharge are considered in addition to the right 
handed neutrino superfield $\hat N^c$. These two doublets $\hat R_{u}$ and 
$\hat R_{d}$ carry non zero R-charges (The R-charge assignments are provided in 
table \ref{R-charges} and therefore, in order to avoid spontaneous R-breaking and 
the emergence of R-axions, the scalar components of $\hat R_{u}$ and $\hat R_{d}$ 
do not receive any nonzero $vev$ and because of this they are coined as inert 
doublets. 

\vspace{-2mm}
\begin{table}[h!]
\begin{center}
\begin{tabular}{|c|ccccccccccccc|}
\hline
\rule{0mm}{5mm}
& $\hat Q_{i}$ & $\hat U_{i}^{c}$ & $\hat D_{i}^{c}$ & $\hat L_{i}$
& $\hat E_{i}^{c}$ & $\hat H_{u}$ & $\hat H_{d}$ & $\hat R_{u}$
& $\hat R_{d}$ & $\hat S$ & $\hat T$ & $\hat O$ & $\hat N^{c}$
\\[0.3em]
\hline
\rule{0mm}{5mm}
$U(1)_{R}$ & 1 & 1 & 1 & 0 & 2 & 0 & 0 & 2 & 2 & 0 & 0 & 0 & 2 \\
[0.3em]
\hline
\end{tabular}
\end{center}
\vspace{-10pt}
\caption{$U(1)_{R}$ charge assignments of the chiral superfields.}
\label{R-charges}
\vspace{-15pt}
\end{table}
\vspace {3mm}

R-symmetry prohibits the gauginos to have Majorana mass term and trilinear scalar 
interactions ($A$-terms) are also absent in a $U(1)_R$ invariant scenario. However, 
the gauginos can acquire Dirac masses. In order to have Dirac gaugino masses one 
needs to include chiral superfields in the adjoint representations of the standard 
model gauge group. Namely a singlet $\hat S$, an $SU(2)_{L}$ triplet $\hat T$ and 
an octet $\hat O$ under $SU(3)_c$. These chiral superfields are essential to 
provide Dirac masses to the bino, wino and gluino respectively. We would like to 
reiterate that the lepton numbers have been identified with the (negative) of 
R-charges such that the lepton number of the SM fermions are the usual ones whereas 
the superpartners of the SM fermions carry {\it non-standard} lepton numbers. 
With such lepton number assignments this R-symmetric model is also lepton number 
conserving \cite{Kumar-1,Claudia,Chakraborty}. 

The generic superpotential carrying an R-charge of two units can be 
written as
\bea
W&=&y^{u}_{ij}\hat H_{u}\hat Q_{i}\hat U^{c}_{j}
+\mu_{u}\hat H_{u}\hat R_{d}+f_{i}\hat L_{i}\hat H_{u}\hat N^{c}
+\lambda_{S}\hat S\hat H_{u}\hat R_{d} +2\lambda_{T}
\hat H_{u}\hat T\hat R_{d}-M_{R}\hat N^{c}\hat S+
\mu_{d}\hat R_{u}\hat H_{d}\nonumber \\
&+&\lambda^\prime_{S}\hat S\hat R_{u}\hat H_{d}
+\lambda_{ijk}
\hat L_{i}\hat L_{j}\hat E^{c}_{k}
+\lambda^{\prime}_{ijk}\hat L_{i}\hat Q_{j}\hat D^{c}_{k}
+2\lambda^\prime_{T}\hat R_{u}\hat T\hat H_{d} 
+y^{d}_{ij}\hat H_{d}\hat Q_{i}\hat D^{c}_{j}+y^{e}_{ij}\hat H_{d}
\hat L_{i}\hat E^{c}_{j} \nonumber \\
&+& \lambda_N {\hat N}^c {\hat H}_u {\hat H}_d.
\label{superpotential}
\eea
For simplicity, in this work we have omitted the terms $\kappa \hat N^c \hat S\hat S$ 
and $\eta \hat N^c$ from the superpotential. As long as $\eta \sim M^2_{\mathrm SUSY}$
and $\kappa \sim 1$ we do not expect any significant change in the analysis and the results 
presented in this paper. 

In order to have a realistic model one should also include supersymmetric 
breaking terms, which are the scalar and the gaugino mass terms. The 
Lagrangian containing the Dirac gaugino masses can be written as 
\cite{Benakli-2,Benakli-1}
\bea
{\cal L}^{\rm Dirac}_{\rm gaugino} &=& \int d^2 \theta 
\dfrac{W^\prime_\alpha}{\Lambda}[\sqrt{2} \kappa_1 ~W_{1 \alpha} {\hat S} 
+ 2\sqrt{2} \kappa_2 ~{\rm tr}(W_{2\alpha} {\hat T}) 
+ 2\sqrt{2} \kappa_3 ~{\rm tr}(W_{3\alpha} {\hat O})] + h.c.,
\label{dirac-gaugino}
\eea
where $W^{\prime}_{\alpha} = \lambda_{\alpha}+\theta_{\alpha}D^{\prime}$ is 
a spurion superfield parametrizing D-type supersymmetry breaking. This results in
Dirac gaugino masses as $D^{\prime}$ acquires $vev$ and are given by
\bea
M^D_i = \kappa_i\frac{<D^{\prime}>}{\Lambda}, 
\eea
where $\Lambda$ denotes the scale of SUSY breaking mediation and $\kappa_i$ are
order one coefficients. 

It is worthwhile to note that these Dirac gaugino mass terms have been dubbed as 
`supersoft' terms. This is because we know that the Majorana gaugino mass terms 
generate logarithmic divergence to the scalar masses  whereas in ref. \cite{Fox}, 
it was shown that the purely scalar loop, obtained from the adjoint superfields 
cancels this logarithmic divergence in the case of Dirac gauginos. Hence it is 
not unnatural to consider the Dirac gaugino masses to be rather large.

The R-conserving but soft supersymmetry breaking terms in the scalar sector are 
generated from a spurion superfield $\hat X$, where $\hat X=x+\theta^2 F_X$ such 
that $R[\hat X]=2$ and $<x>=0$, $<F_X>\neq 0$. The non-zero $vev$ of $F_X$ 
generates the scalar soft terms and the corresponding potential is given by 
\bea
V_{soft}&=& m^{2}_{H_{u}} H_{u}^{\dagger}H_{u}+m^{2}_{R_{u}}
R_{u}^{\dagger}R_{u}+ m^{2}_{H_{d}}H_{d}^{\dagger}H_{d}
+m^{2}_{R_{d}}R_{d}^{\dagger}R_{d}
+m^{2}_{\tilde L_{i}}\tilde L_{i}^{\dagger}\tilde L_{i}+
m^2_{{\tilde R}_i}{{\tilde l}^\dagger_{Ri} {\tilde l}_{Ri}} \nonumber \\
&+& M_{N}^{2}\tilde N^{c\dagger}\tilde N^{c}
+m_{S}^{2} S^{\dagger}S+2m_{T}^{2} {\rm tr}(T^{\dagger}T)
+2m_O^2 {\rm tr}(O^\dagger O) 
+ (B\mu H_u H_d + {\rm h.c.}) \nonumber \\
&-& (b\mu_L^i H_u {\tilde L}_i + {\rm h.c.}) 
+(t_{S}S+{\rm h.c.})+\frac{1}{2} b_{S}(S^{2}+{\rm h.c.})
+b_{T} ({\rm tr}(TT) + {\rm h.c.})\nonumber \\ 
&+&B_O({\rm tr}(OO) + {\rm h.c.}).
\label{soft-scalar-terms}
\eea
The presence of the bilinear terms $b\mu_L^i H_u {\tilde L}_i$ implies
that all the three left handed sneutrinos can acquire non-zero $vev$'s. 
However, it is always possible to make a basis rotation in which only one
of the left handed sneutrinos get a non-zero $vev$ and one must keep in mind
that the physics is independent of this basis choice. 

Such a rotation can be defined as 
\bea
\hat L_{i}=\frac{v_{i}}{v_{a}}\hat L_{a} + \sum_b {e_{ib}\hat L_{b}}.
\eea
Note that the index $(i)$ runs over three generations whereas $a = 1(e)$ and 
$b = 2,3 (\mu, \tau)$. This basis rotation implies that the scalar component of 
the superfield $\hat L_a$ acquires a non zero $vev$ (i.e. $\langle {\tilde \nu} 
\rangle \equiv v_a \neq 0)$ whereas the other two sneutrinos do not get any 
$vev$. One can further go to a basis where the charged lepton Yukawa 
couplings are diagonal. It is, however, important to note that the charged lepton 
of flavor $a$ (i.e. the electron) cannot get mass from this Yukawa couplings 
because of $SU(2)_L$ invariance but can be generated from R-symmetric supersymmetry 
breaking operators \cite{Kumar-1}. Moreover, we also choose the neutrino Yukawa 
coupling in such a way that only $\hat L_a$ couples to\footnote{For a detailed 
discussion we refer the reader to ref. \cite{Chakraborty}.} ${\hat N}^c$. In such 
a scenario the left-handed sneutrino can play the role of a down type Higgs boson 
since its $vev$ preserves lepton number and is not constrained by neutrino Majorana 
mass. Hence one has the freedom to keep a very large $\mu_d$ such that the 
superfields $\hat H_d$ and $\hat R_u$ get decoupled from the theory. This is what 
we shall consider in the rest of our discussion.

With a single sneutrino acquiring a $vev$ and in the mass eigenstate basis of the 
charged lepton and down type quark fields the superpotential now has the following 
form (integrating out ${\hat H}_d$ and ${\hat R}_u$)
\bea
W&=&y_{ij}^{u}\hat H_{u}\hat Q_{i}\hat U_{j}^{c}+\mu_{u}\hat 
H_{u}\hat R_{d}+f\hat L_{a}\hat H_{u}\hat N^{c}+
\lambda_{S}\hat S\hat H_{u}\hat R_{d} 
+2\lambda_{T}\hat H_{u}
\hat T\hat R_{d} \nonumber \\
&-&M_{R}\hat N^{c}\hat S + W^{\prime},
\label{final-superpotential}
\eea
where
\bea
&&W^{\prime}=\sum_{b=2,3} f^l_b {\hat L}_a {\hat L^\prime}_b 
{\hat E^{\prime c}}_b + \sum_{k=1,2,3} f^d_k {\hat L}_a
{\hat Q^\prime}_k {\hat D^{\prime c}}_k \nonumber \\
&&+ \sum_{k=1,2,3} \dfrac{1}{2} 
{\tilde \lambda}_{23k}{\hat L^\prime}_2 {\hat L^\prime}_3 
{\hat E^{\prime c}}_k + \sum_{j,k=1,2,3;b=2,3}{\tilde \lambda}^\prime_{bjk}
{\hat L^\prime}_b {\hat Q^\prime}_j {\hat D^{\prime c}}_k, \nonumber \\
\label{W-diag}
\eea
and includes all the trilinear R-parity violating terms in this model.
In the subsequent discussion we shall confine ourselves to this choice of basis
but get rid of the primes from the fields and make the replacement ${\tilde 
\lambda}$, ${\tilde \lambda}^\prime$ $\rightarrow$ $\lambda$,$\lambda^\prime$.

In this rotated basis the soft supersymmetry breaking terms look like
\bea
V_{soft}&=& m^{2}_{H_{u}}H_{u}^{\dagger}H_{u}+m^{2}_{R_{d}}
R_{d}^{\dagger}R_{d}+m^{2}_{\tilde L_{a}} \tilde L_{a}^{\dagger}
\tilde L_{a} 
+\sum_{b=2,3} m^{2}_{\tilde L_{b}} \tilde L_{b}^{\dagger}
{\tilde L_{b}}+M_{N}^{2}{\tilde N}^{c\dagger} {\tilde N}^{c}
+m^2_{{\tilde R}_i}{{\tilde l}^\dagger_{Ri} {\tilde l}_{Ri}} \nonumber \\
&+&
+m_{S}^{2} S^{\dagger}S+2m_{T}^{2} {\rm tr}(T^{\dagger}T) 
+2m_O^2 {\rm tr}(O^\dagger O) 
- (b\mu_L H_u {\tilde L}_a + {\rm h.c.}) +(t_{S}S+{\rm h.c.})\nonumber \\
&+&\frac{1}{2} b_{S}(S^{2}+{\rm h.c.})
+b_{T} ({\rm tr}(TT) + {\rm h.c.}) +B_O({\rm tr}(OO) + {\rm h.c.}).
\label{final-softsusy-terms} 
\eea
With this short description of the theoretical framework let us now explore the 
scalar and the fermionic sectors in some detail in order to prepare the ground 
for the study of the diphoton decay of the lightest Higgs boson.
\section{The scalar sector}
The scalar potential receives contributions from the F-term, the D-term, the soft 
SUSY breaking terms and the terms coming from one-loop radiative corrections. 
Thus, schematically,
\bea
V = V_{\rm F} + V_{\rm D} + V_{\rm soft} + V_{\rm one-loop}.
\eea
The F-term contribution is given by
\bea
V_{F} &=& \sum_i \left|\frac{\partial W}{\partial \phi_i}\right|^2,
\eea
where the superpotential $W$ is given by eq.~(\ref{final-superpotential}). The 
$D$-term contribution can be written as
\bea
\label{gauge}
V_{D}=\frac{1}{2}\sum_{a}D^{a}D^{a}+\frac{1}{2}D_{Y}D_{Y},
\eea
where
\bea
\label{aux}
D^{a}&=&g(H_{u}^{\dagger}\tau^{a}H_{u}+\tilde L_{i}^{\dagger}
\tau^{a}\tilde L_{i}+T^{\dagger}\lambda^{a} T) 
+ \sqrt 2 (M_{2}^{D}T^{a}+M_{2}^{D}T^{a\dagger}).
\eea
The $\tau^{a}$'s and $\lambda^{a}$'s are the $SU(2)$ generators in the fundamental 
and adjoint representation respectively. The weak hypercharge contribution $D_{Y}$ 
is given by
\bea
\label{aux-1}
D_{Y}=\frac{g^{\prime}}{2}(H_{u}^{+}H_{u}-\tilde L_{i}^{+}\tilde L_{i})
+\sqrt 2 M_{1}^{D}(S+S^{\dagger}),
\eea
where $g$ and $g^\prime$ are $SU(2)_L$ and $U(1)_Y$ gauge couplings respectively.
The expanded forms of $V_F$ and $V_D$ in terms of various scalar fields can be 
found in \cite{Chakraborty}. The soft SUSY breaking term $V_{\rm soft}$ is given 
in Eq.~(\ref{final-softsusy-terms}) whereas the dominant radiative corrections to 
the quartic potential are of the form $\frac{1}{2}\delta\lambda_{u} (|H_{u}|^2)^2$, 
$\frac{1}{2}\delta\lambda_{\nu}(|\tilde\nu_{a}|^2)^2$ and $\frac{1}{2} 
\delta\lambda_{3}|H_{u}^{0}|^2|\tilde \nu_{a}|^2$. The coefficients $\delta
\lambda_{u}$, $\delta\lambda_{\nu}$ and $\delta\lambda_{3}$ are given by
\bea
\delta\lambda_{u}&=& \frac{3 y_{t}^{4}}{16\pi^{2}}
\ln \left(\frac{m_{\tilde t_{1}}m_{\tilde t_{2}}}{m_{t}^2}\right)
+\frac{5\lambda_{T}^{4}}{16\pi^{2}}\ln\left(\frac{m_{T}^{2}}
{v^{2}}\right) \nonumber 
+\frac{\lambda_{S}^{4}}{16\pi^{2}}\ln\left(\frac{m_{S}^2}i
{v^{2}}\right)\nonumber \\
&-&\frac{1}{16\pi^{2}}\frac{\lambda_{S}^{2}\lambda_{T}^{2}}
{m_{T}^{2}-m_{S}^{2}}
\left(m_{T}^{2}\bigg\{\ln\left(\frac{m_{T}^{2}}{v^{2}}\right)-1\bigg\}
-m_{S}^{2}\bigg\{\ln \left(\frac{m_{S}^{2}}{v^{2}}\right)-1\bigg\}
\right),\nonumber \\
\eea
\bea
\delta\lambda_{\nu}&=& \frac{3 y_{b}^{4}}{16\pi^{2}}
\ln \left(\frac{m_{\tilde b_{1}}m_{\tilde b_{2}}}{m_{b}^2}\right)
+\frac{5\lambda_{T}^{4}}{16\pi^{2}}\ln\left(\frac{m_{T}^{2}}
{v^{2}}\right)
+\frac{\lambda_{S}^{4}}{16\pi^{2}}
\ln\left(\frac{m_{S}^2}{v^{2}}\right)\nonumber \\
&-& \frac{1}{16\pi^{2}}\frac{\lambda_{S}^{2}\lambda_{T}^{2}}
{m_{T}^{2}-m_{S}^{2}}
\left(m_{T}^{2}\bigg\{\ln\left(\frac{m_{T}^{2}}{v^{2}}\right)-1\bigg\}
-m_{S}^{2}\bigg\{\ln \left(\frac{m_{S}^{2}}{v^{2}}\right)-1\bigg\}
\right), \nonumber \\
\eea
\bea
\delta\lambda_{3}&=& \frac{5 \lambda_{T}^{4}}{32\pi^{2}}
\ln (\frac{m_{T}^{2}}{v^{2}})
+\frac{1}{32\pi^{2}}\lambda_{S}^{4}
\ln\left(\frac{m_{S}^{2}}{v^{2}}\right)
+\frac{1}{32\pi^{2}}\frac{\lambda_{S}^{2}\lambda_{T}^{2}}{m_{T}^{2}
-m_{S}^{2}}\bigg(m_{T}^{2}\bigg\{\ln \left(\frac{m_{T}^{2}}
{v^{2}}\right)-1\bigg\}\nonumber \\
&-&m_{S}^{2}\bigg\{\ln\left(\frac{m_{S}^{2}}{v^{2}}\right)-1\bigg\}\bigg).
\eea
We shall see later that for large values of the couplings $\lambda_T$ and 
$\lambda_S$ or large stop masses these one-loop radiative contributions to the
Higgs quartic couplings could play important roles in obtaining a CP-even
lightest Higgs boson with a mass around 125 GeV. 
\vspace*{-0.25in}
\subsection{CP-even neutral scalar sector}
\label{cp-even-neutral-scalar}
Let us assume that the neutral scalar fields $H^0_u$, ${\tilde \nu}_a ~(a = 1(e))$, 
$S$ and $T$ acquire real vacuum expectation values $v_u$, $v_a$, $v_S$ and $v_T$, 
respectively. The scalar fields $R_d$ and ${\tilde N}^c$ carrying R-charge 2 are 
decoupled from these four scalar fields. We can split the fields in terms of their 
real and imaginary parts: $H_{u}^{0}=h_{R} +i h_{I}$, ${\tilde \nu}^a={\tilde 
\nu}^a_{R}+i{\tilde \nu}^a_{I}$, $S=S_{R}+iS_{I}$ and $T=T_{R}+iT_{I}$. The 
resulting minimization equations can be found easily and with the help of 
these minimization equations, the neutral CP-even scalar squared-mass matrix in 
the basis $(h_{R},\tilde\nu_{R},S_{R},T_{R})$ can be written down in a 
straightforward way, where $h_4$ corresponds to the lightest CP even 
mass eigenstate \cite{Chakraborty}. In the R-symmetry preserving scenario the elements 
of this symmetric $4\times 4$ matrix are found to be
\bea
\label{CP-even}
(M_S^2)_{11}&=&\frac{(g^{2}+g^{\prime 2})}{2}v^{2}\sin^{2}\beta+
(fM_{R}v_{S}-b\mu_{L}^{a})(\tan\beta)^{-1} 
+2\delta\lambda_{u}v^{2}\sin^{2}\beta,\nonumber \\
(M_S^2)_{12}&=&f^{2}v^{2}\sin2\beta+b\mu_{L}^{a}-
\frac{(g^{2}+g^{\prime 2}-2\delta\lambda_{3})}{4}v^{2}
\sin2\beta 
- fM_{R}v_{S},\nonumber \\
(M_S^2)_{13}&=& 2\lambda_{S}^{2}v_{S}v\sin\beta+2\mu_{u}
\lambda_{S}v\sin\beta+2\lambda_{S}\lambda_{T}v v_{T}\sin\beta 
+\sqrt2 g^{\prime}M_{1}^{D}v\sin\beta-fM_{R}v\cos\beta,\nonumber \\
(M_S^2)_{14}&=& 2\lambda_{T}^{2}v_{T}v\sin\beta+2\mu_{u}\lambda_{T}
v\sin\beta+2\lambda_{S}\lambda_{T}v_{S}v\sin\beta 
- \sqrt 2 gM_{2}^{D} v\sin\beta,\nonumber \\
(M_S^2)_{22}&=&\frac{(g^{2}+g^{\prime 2})}{2}v^{2}\cos^{2}\beta+
(fM_{R}v_{S}-b\mu_{L}^{a})\tan\beta 
+ 2\delta\lambda_{\nu}v^{2}\cos^{2}\beta,\nonumber \\
(M_S^2)_{23}&=& -\sqrt 2 g^{\prime}M_{1}^{D}v\cos\beta
-fM_{R}v\sin\beta,\nonumber \\
(M_S^2)_{24}&=&\sqrt 2 g M_{2}^{D} v\cos\beta,\nonumber \\
(M_S^2)_{33}&=&-\mu_{u}\lambda_{S}\frac{v^{2}\sin^2\beta}{v_{S}}
-\frac{\lambda_{S}\lambda_{T}v_{T}v^{2}\sin^2\beta}{v_{S}} 
-\frac{t_S}{v_{S}}+\frac{g^{\prime}M_{1}^{D}v^{2}\cos2\beta}{\sqrt 2 v_{S}}
+\frac{f M_{R}v^{2}\sin2\beta}{2v_{S}},\nonumber \\
(M_S^2)_{34}&=&\lambda_{S}\lambda_{T}v^{2}\sin^2\beta, \nonumber \\
(M_S^2)_{44}&=&-\mu_{u}\lambda_{T}\frac{v^{2}}{v_{T}}\sin^2\beta
-\lambda_{S}\lambda_{T}v_{S}\frac{v^{2}}{v_{T}}\sin^2\beta 
-\frac{g M_{2}^{D}}{\sqrt 2} \frac{v^{2}}{v_{T}}\cos2\beta,
\eea
where $\tan\beta = v_u/v_a$ and $v^2 = v^2_u + v^2_a$. 
The $W^\pm$- and the $Z$-boson masses can be written as
\bea
m^2_W = \dfrac{1}{2}g^2 (v^2 + 4 v_T^2), \nonumber \\
m^2_Z = \dfrac{1}{2} g^2 v^2/\cos^2{\theta_W}.
\eea

Note that the electroweak precision measurements of the $\rho$-parameter requires 
that the triplet $vev$ $v_T$ must be small ($\lsim$ 3 GeV) \cite{PDG}. In addition, 
our requirement of a doublet-like lightest CP-even Higgs boson, in turn, demands a 
small $vev$ $v_S$ of the singlet $S$ as well. This is because a small value of $v_S$ 
reduces the mixing between the doublets and the singlet scalar $S$. In such a 
simplified but viable scenario in which the singlet and the $SU(2)_L$ triplet 
scalars get decoupled from the theory, we are left with a $2\times 2$ scalar mass 
matrix. In this case the angle $\alpha$ represents the mixing angle between $h_R$ 
and $\tilde \nu_R$ and can be expressed in terms of other parameters as follows
\bea
\tan2\alpha&=& -2 \frac{f^2 v^2 \sin 2\beta + b\mu_L^a - 
\frac{(g^2 + g^{\prime 2}-2\delta\lambda_3)}{4} v^2 \sin2\beta}
{\frac{(g^2 + g^{\prime 2}) v^2 \cos 2\beta}{2} + 2b\mu_L^a \cot 2\beta
-2 v^2 \big\{\delta\lambda_u \sin^2 \beta - \delta\lambda_{\nu} \cos^2 \beta\big\}}.
\label{mixing-angle-alpha}
\eea  
\subsection{Tree level mass bound on $m_h$}
\label{tree-level-mass-bound}

In addition, in such a situation (with $v_S, ~v_T \ll v$) it can be shown easily 
that the lightest CP-even Higgs boson mass is bounded from above at tree level 
\cite{Chakraborty}, 
\bea
(m_{h}^{2})_{\rm tree} \leq m_z^2 \cos^2 2\beta + f^2 v^2 \sin^2 2\beta.
\label{tree-level-higgs-mass-bound}
\eea
\begin{figure}[htb]
\begin{center}
\includegraphics[height=3in,width=4in]{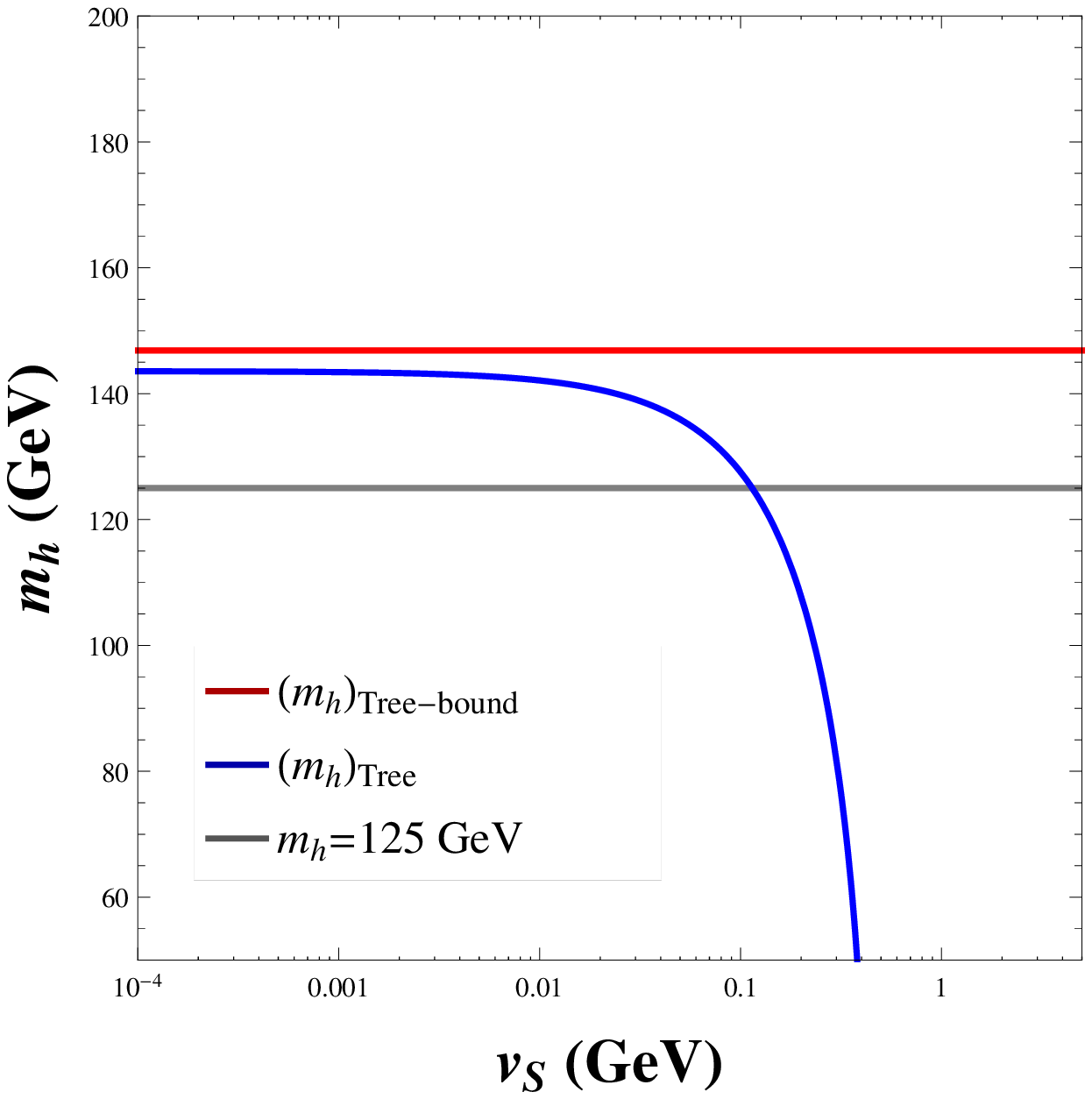} 
\caption{\label{higgs_mass_vs} The tree level mass of the lightest Higgs boson as 
a function of the singlet ($S$) vacuum expectation value $v_S$ with $f$ = 1.5, 
$\tan\beta$ =4 and other parameter choices are as described in the text. 
The upper bound on the tree level mass of the Higgs boson from 
eq.~\ref{tree-level-higgs-mass-bound} is also shown.} 
\end{center}
\end{figure}
\begin{figure}[htb]
\begin{center}
\includegraphics[height=3in,width=4in]{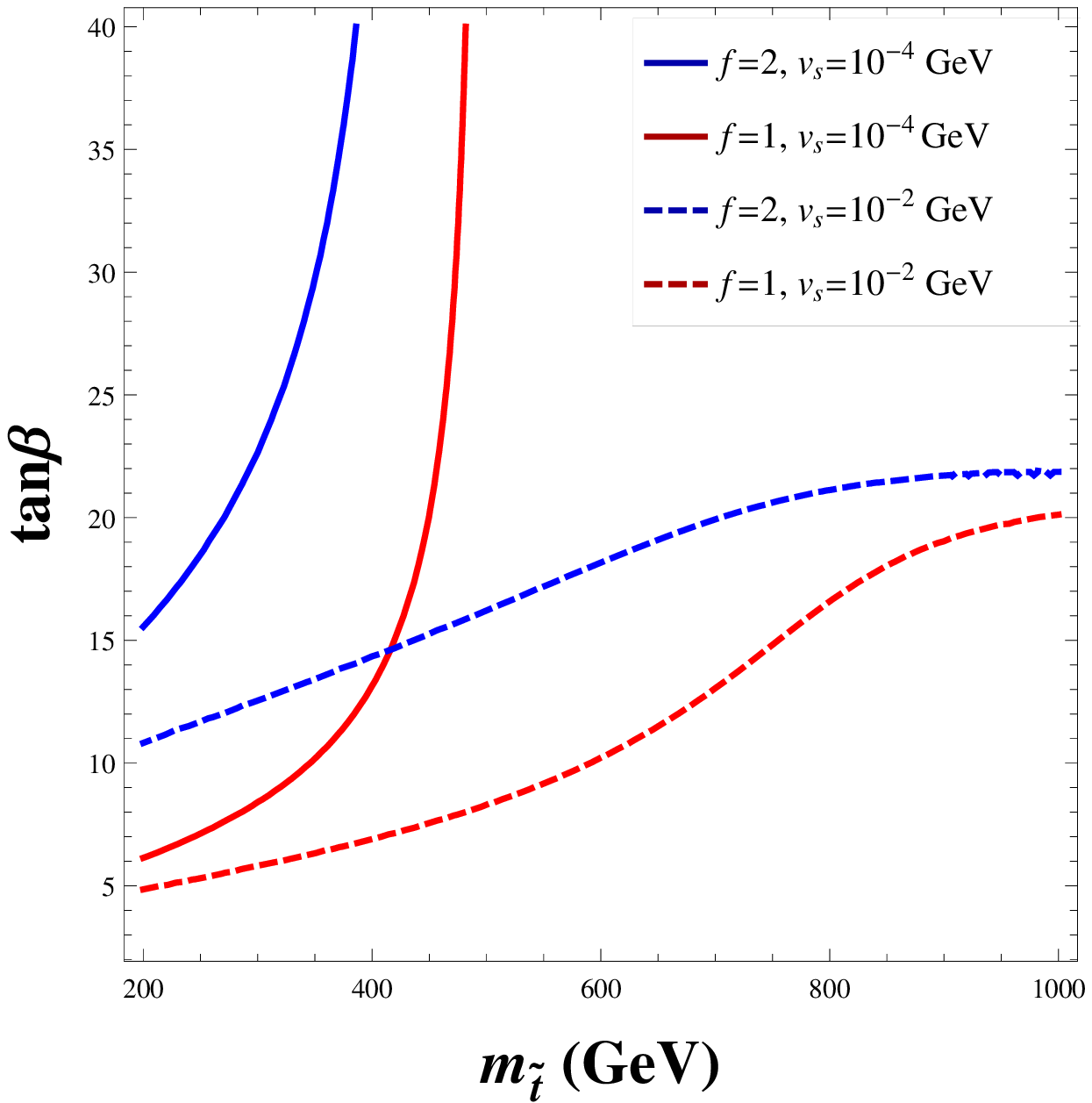} 
\caption{\label{higgs_stop_tanb} Mass-contours for the lightest Higgs boson with
$m_h = $ 125 GeV in the $m_{\tilde t}$ --$\tan\beta$ plane for large values 
of $f$ and $\lambda_T$ = 0.5.} 
\end{center}
\end{figure}
The bound in Eq.~(\ref{tree-level-higgs-mass-bound}) is saturated for $v_s \lsim 
10^{-3}$ GeV, i.e., when the singlet has a large soft supersymmetry breaking mass 
and is integrated out. The $f^2 v^2$ term grows at small $\tan\beta$ and thus the 
largest Higgs boson mass is obtained with low $\tan\beta$ and large values of $f$. 
We shall show in the next section that $f \sim 1$ can be accommodated in this 
scenario without spoiling the smallness of the neutrino mass at tree level. Therefore, 
for $f\sim \mathcal O(1)$, the tree level Higgs boson mass can be as large as  
$\sim 125$ GeV where the peak in the diphoton invariant mass has been observed and 
no radiative corrections are required. This means that in this scenario one can 
still  afford a stop mass as small as 350 GeV or so and couplings $\lambda_T$ and 
$\lambda_S$ can be small ($\sim 10^{-4}$) as well. This is illustrated in figure 
(\ref{higgs_mass_vs}) where, the lightest Higgs boson mass is shown as a function 
of $v_S$ for $f$ = 1.5, $\tan\beta$ = 4 and for a set of other parameter choices 
discussed later. One can see that for a very small $v_S$ ($\lsim 10^{-3}$ GeV) the 
tree level Higgs boson mass is 150 GeV and is reduced to 125 GeV for a $v_S 
\sim 0.2$ GeV. As $v_S$ increases further, $(M_h)_{\rm Tree}$ starts decreasing 
rapidly and the Higgs boson mass becomes lighter than 100 GeV. In  such a case one 
requires larger radiative corrections to the Higgs boson mass and this can be 
achieved with the help of large triplet/singlet couplings ($\mathcal O(1)$)
and/or large stop mass.  For example, with a choice of $\lambda_S = 0.91$ 
and $\lambda_T$ = 0.5, the one-loop radiative corrections to the Higgs boson mass 
arising from these two couplings are sizable\footnote{These choices of $\lambda_T$ 
and $\lambda_S$ are not completely independent. Rather they follow a relationship derived 
from the requirement of small tree level mass of the active neutrino. This will be 
discussed in the next section.}. In this case, in order to have a 125 GeV Higgs 
boson, the tree level contribution should be smaller and for a very small 
$v_S$ ($\sim 10^{-4}$ GeV) and large $f$ ($\gsim$ 1), this can be achieved with a 
larger $\tan\beta$. The one loop corrections from the stop loop 
must also be small and this is realized for small $m_{\tilde t}$ and large 
$\tan\beta$. This is illustrated in figure (\ref{higgs_stop_tanb}) where we 
plot mass-contours for the lightest Higgs boson with a mass of 125 GeV in the  
$m_{\tilde t}$ -- $\tan\beta$ plane for different choices of $f$ and $v_S$. One 
can see from this figure the effect of a larger $v_S$, which requires a larger 
stop loop contribution to have a Higgs boson mass of 125 GeV.  

\section{The fermionic sector}
The fermionic sector of the scenario, involving the neutralinos and the 
charginos, has rich new features. In the context of the present 
study, when analyzed in conjunction with the scalar sector of the scenario,
this sector plays a pivotal role by presenting the defining issues for the 
phenomenology of this
scenario. Its influence ranges over physics of the Higgs boson at current 
experiments and the physics of the neutrinos before finally reaching out to 
the domain of
astrophysics and cosmology by offering a possible warm dark matter candidate
whose actual presence may find support in the recent observations of a
satellite-borne X-ray
experiment. Thus, it is of crucial importance to study the structure
and the content of this sector in appropriate detail. 

A thorough discussion of $\mu_{\gamma\gamma}$ in the present scenario 
requires a study of the masses and the mixing angles of the neutralinos and
the charginos. One of the natural consequences of such a $U(1)_R$-lepton 
number model with a right-handed neutrino is that one of the left-handed
neutrinos (the electron-type one) and the right-handed neutrino become
parts of the extended neutralino mass matrix. The electron-type neutrino of 
the SM can be identified with the lightest neutralino eigenstate. We also 
address the issue of tree level neutrino mass. Subsequently, we show that 
in certain region of the parameter space the lightest neutralino-like
state can be very light (with a mass of order 100 MeV). This may contribute 
to the invisible decay width of the lightest Higgs boson. We study the 
validity of the parameter space when it is subject to the constraint from
invisible decay width of the Higgs boson.

\subsection{The neutralino sector: R-conserving case}
\label{neutralino-r-conserving}
In the neutral fermion sector we have mixing among the Dirac gauginos, the 
higgsinos, the active neutrino of flavor $`a$' (i.e., $\nu_e$) and the single 
right-handed neutrino $N^c$ once the electroweak symmetry is broken. The part 
of the Lagrangian that corresponds to the neutral fermion mass matrix is given 
by $\mathcal{L}=(\psi^{0+})^{T}M_{\chi}^D(\psi^{0-})$ where $\psi^{0+}
=(\tilde b^{0},\tilde w^{0},\tilde R_{d}^{0},N^{c})$, with R-charges $+1$ 
and $\psi^{0-}=(\tilde S,\tilde T^{0},\tilde H_{u}^{0},\nu_{e})$ with R-charges 
-1. The neutral fermion mass matrix $M_{\chi}^D$ is given by
\bea
M_{\chi}^D=\left(
\begin{array}{cccc}
M_{1}^{D} & 0 & \frac{g^{\prime}v_{u}}{\sqrt 2}& -\frac{g^{\prime}v_{a}}{\sqrt 2}\\
0 & M_{2}^{D} & -\frac{gv_{u}}{\sqrt 2}& \frac{gv_{a}}{\sqrt 2}\\
\lambda_{S}v_{u} & \lambda_{T}v_{u} & \mu_{u}+\lambda_{S}v_{S}
+\lambda_{T}v_{T} & 0\\
M_{R} & 0 & -fv_{a} & -fv_{u}
\end{array} \right).
\label{neutrino-neutralino-matrix}
\eea
The above matrix can be diagonalized by a biunitary transformation involving two 
unitary matrices $V^N$ and $U^N$ and results in four Dirac mass eigenstates 
${\widetilde \chi}^{0+}_i \equiv \left( \begin{array}{c} {\widetilde \psi}^{0+}_i \\
\overline {{\widetilde \psi}^{0-}_i}\end{array} \right)$, with $i=1,2,3,4$ and 
${\widetilde \psi}^{0+}_i = V^N_{ij} \psi^{0+}_j$, ${\widetilde \psi}^{0-}_i = 
U^N_{ij} \psi^{0-}_j$. The lightest mass eigenstate ${\widetilde \chi}^{0+}_4$ 
is identified with the light Dirac neutrino. The other two active neutrinos 
remain massless in this case. Generically the Dirac neutrino mass can be in the 
range of a few eV to tens of MeV. However, one can also accommodate a mass of 
0.1 eV or smaller for the Dirac neutrino by assuming certain relationships 
involving different parameters \cite{Chakraborty}, which are  
\bea
\lambda_{T}&=& \lambda_{S} \tan\theta_{W}
\label{relation-lambda-s-t}
\eea
and 
\bea
M_{R}&=&\frac{\sqrt 2f M_{1}^{D}\tan\beta}{g\tan\theta_{W}}.
\label{relation-MR}
\eea
With these choices the Dirac mass of the neutrino can be written as
\bea
m^D_{\nu_e}&=& \frac{v^3 f g \sin\beta}{\sqrt 2 \gamma M_1^D M_2^D}
\lambda_T (M_2^D-M_1^D).
\label{nu-dirac-mass}
\eea
where $\gamma=\mu_u+\lambda_S v_S+\lambda_T v_T$. It is straightforward to check 
from eq.~(\ref{nu-dirac-mass}) that by suitable choices of the parameters $f$, 
$\lambda_T$ and $\epsilon \equiv (M_2^D - M_1^D)$, one can have a Dirac neutrino 
mass in the right ballpark of $\lsim$ 0.1 eV. Note that a choice of large $f 
\sim {\mathcal O}(1)$ is possible for a small $\lambda_T$ ($\sim 10^{-6}$) and 
nearly degenerate Dirac gauginos ($\epsilon \lsim 10^{-1}$) assuming $\mu_u$, 
$M^D_2$, $M^D_1$ in the few hundred GeV range.
%
\subsection{The neutralino sector: R-breaking case}
\label{sec:neutralino-r-breaking}
R-symmetry is not an exact symmetry and is broken by a small gravitino mass. One 
can therefore consider the gravitino mass as the order parameter of R-breaking. 
The breaking of R-symmetry has to be communicated to the visible sector and in 
this work we consider anomaly mediation of supersymmetry breaking playing the role 
of the messenger of R-breaking. This is known as anomaly mediated R-breaking (AMRB) 
\cite{Bertuzzo}. A non-zero gravitino mass generates Majorana gaugino masses and 
trilinear scalar couplings. We shall consider the R-breaking effects to be small 
thus limiting the gravitino mass ($m_{3/2}$) around 10 GeV. 

The R-breaking Lagrangian contains the following terms
\bea
\mathcal{L}&=& M_{1}\widetilde b^0 \widetilde b^0 +M_{2}\widetilde 
w^0 \widetilde w^0 +M_{3}\widetilde g\widetilde g +
\sum_{b=2,3} A^l_b {\tilde L}_a {\tilde L}_b {\tilde E^{c}}_b 
+ \sum_{k=1,2,3} A^d_k {\tilde L}_a
{\tilde Q}_k {\tilde D^{c}}_k + \sum_{k=1,2,3} \dfrac{1}{2} 
A^{\lambda}_{23k}{\tilde L}_2 {\tilde L}_3
{\tilde E^{c}}_k \nonumber \\
&+& \sum_{j,k=1,2,3;b=2,3} A^{{\lambda^\prime}}_{bjk}{\tilde L}
_b {\tilde Q}_j {\tilde D^{c}}_k
+A^{\nu}H_{u}\tilde L_{a}\tilde N^{c}
+H_{u} \tilde Q A^u \tilde U^{c}
\eea 
where $M_1$, $M_2$ and $M_3$ are the Majorana mass parameters corresponding to 
$U(1)$, $SU(2)$ and $SU(3)$ gauginos, respectively and $A$'s are the scalar 
trilinear couplings.

The (Majorana) neutralino mass matrix containing R-breaking effects can
be written in the basis $\psi^0 = (\tilde b^0, \tilde S, \tilde w^0, \tilde T,
\tilde R_d, \tilde H_u^0, N_c, \nu_e)^T$ as
\bea
{\cal L}^{\rm mass}_{{\tilde \chi}^0} = \dfrac{1}{2} (\psi^0)^T M_{\chi}^{M} 
\psi^0 + h.c. 
\eea
where the symmetric ($8 \times 8$) neutralino mass matrix $M_{\chi}^{M}$ is given 
by 
\bea
M_{\chi}^{M}=\left(
\begin{array}{cccccccc}
M_{1} & M_{1}^{D} & 0 & 0 & 0 & \frac{g^{\prime}v_{u}}{\sqrt 2} & 0
& -\frac{g^{\prime}v_{a}}{\sqrt 2}\\
M_{1}^{D} & 0 & 0 & 0 & \lambda_{S}v_{u} & 0 & M_{R} & 0\\
0 & 0 & M_{2} & M_{2}^{D} & 0 & -\frac{g v_{u}}{\sqrt 2} & 0 &
\frac{g v_{a}}{\sqrt 2}\\
0 & 0 & M_{2}^{D} & 0 & \lambda_{T}v_{u} & 0 & 0 & 0 \\
0 & \lambda_{S}v_{u} & 0 & \lambda_{T}v_{u} & 0 & \mu_{u}+
\lambda_{S}v_{S}+\lambda_{T}v_{T} & 0 & 0\\
\frac{g^{\prime}v_{u}}{\sqrt 2} & 0 & -\frac{gv_{u}}{\sqrt 2}& 0 &
\mu_{u}+\lambda_{S}v_{S}+\lambda_{T}v_{T} & 0 & -fv_{a} & 0\\
0 & M_{R} & 0 & 0 & 0 & -fv_{a} & 0 & -fv_{u} \\
-\frac{g^{\prime}v_{a}}{\sqrt 2}& 0 & \frac{g v_{a}}{\sqrt 2} & 0
& 0 & 0 & -fv_{u} & 0
\end{array} \right).\nonumber \\
\label{majorana-neutralino}
\eea
The above mass matrix can be diagonalized by a unitary transformation given by
\bea
N^\star M_{\chi}^{M} N^\dagger = (M_{\chi})_{\rm diag}.
\eea
The two-component mass eigenstates are defined by
\bea
\chi^0_i = N_{ij} \psi^0_j, ~~~~~~i,j = 1,...,8
\eea
and one can arrange them in Majorana spinors defined by
\bea
{\tilde \chi}^0_i = \left(
\begin{array}{c}
\chi^0_i \\
{\bar \chi}^0_i
\end{array} \right), ~~~~~~i = 1,...8.
\eea
Similar to the Dirac case, the lightest eigenvalue ($m_{{\tilde \chi}^0_8}$) of 
this neutralino mass matrix corresponds to the Majorana neutrino mass. Using the 
expression of $M_R$ in eq.~(\ref{relation-MR}) and the relation between $\lambda_S$ 
and $\lambda_T$ in eq.~(\ref{relation-lambda-s-t}), the active neutrino mass is 
given by~\cite{Chakraborty},
\bea
(m_\nu)_{\rm Tree} &=&-v^{2}\frac{\left[g \lambda_{T} v^{2}
(M_{2}^{D}-M_{1}^{D})\sin\beta\right]^{2}}{\left[M_{1}\alpha^{2}
+M_{2}\delta^{2}\right]}
\label{neutrino_majorana}
\eea
where $\alpha$ and $\delta$ are defined as
\bea
\alpha&=&\frac{2 M_{1}^{D} M_{2}^{D}\gamma\tan\beta}
{g\tan\theta_{w}}
+\sqrt 2  v^{2}\lambda_{S}\tan\beta(M_{1}^{D}\sin^{2}\beta+
M_{2}^{D}\cos^{2}\beta),\nonumber \\
\delta&=&\sqrt2  M_{1}^{D}v^{2}\lambda_{T}\tan\beta
\eea
and the quantity $\gamma$ has been defined earlier in section 
\ref{neutralino-r-conserving}.
This shows that to have an appropriate neutrino mass we require the Dirac gaugino 
masses to be highly degenerate. The requirement on the degree of degeneracy can be 
somewhat relaxed if one chooses an appropriately small value of $\lambda_T$. Such 
a choice, in turn, would imply an almost negligible radiative contribution to the 
lightest Higgs boson mass. Interestingly, the Yukawa coupling does not appear in 
the expression for $(m_\nu)_{\rm Tree}$ in eq.~(\ref{neutrino_majorana}). This is 
precisely because of the relation between $M_R$ and $f$ in eq.~(\ref{relation-MR}). 
However, `$f$' has some interesting effects on the next-to-lightest eigenstates of 
the mass matrix. The following situations are phenomenologically important:
%
\begin{itemize}
\item A large value of $f\sim \mathcal O(1)$ generates a very light bino-like 
neutralino (${\tilde \chi}^0_7$) with mass around a few hundred MeV. In this case, 
this is the lightest supersymmetric particle (LSP) and its mass is mainly controlled by 
the R-breaking Majorana gaugino mass parameter $M_1$. A very light neutralino has 
profound consequences in both cosmology as well as in collider physics 
\cite{Dreiner,Hooper,Lebedev,Ota,Kim,Dreiner:2009er,Choudhury,Adhikari,Adhikari-1,Adhikari-2}. 
In the context of the present model one can easily satisfy the stringent 
constraint coming from the invisible decay width of the $Z$ boson because 
the light neutralino is predominantly a bino. One should also take into 
account the constraints coming from the invisible decay branching ratio 
of the lightest Higgs boson. In our scenario $h\rightarrow {\tilde \chi}^0_7 {\tilde \chi}^0_8$
(where ${\tilde \chi}^0_8$ is the light active neutrino) could effectively contribute to the 
invisible final state. This is because, although ${\tilde \chi}^0_7$ would undergo an 
R-parity violating decay, for example, ${\tilde \chi}^0_7 \rightarrow e^+ e^- \nu$, the 
resulting four body final state presumably has to be dealt with as an invisible mode
for the lightest Higgs boson. Such constraints are discussed in detail later in this paper. 
Note that $\Gamma(h\rightarrow {\tilde \chi}^0_7 {\tilde \chi}^0_7)$ is negligibly small 
because of suppressed $h$-${\tilde \chi}^0_7$-${\tilde \chi}^0_7$ coupling for a bino-dominated, 
${\tilde \chi^0_7}$.

A 10 GeV gravitino NLSP could also decay to a final state comprising of the lightest neutralino accompanied
by a photon. In order to avoid the strong constraint on such a decay process coming from big-bang nucleosynthesis
(BBN) one must consider an upper bound on the reheating temperature of the universe $T_R \lsim 10^6$ GeV \cite{Kim,Moroi}.
In addition, such a light state is subjected to various collider bounds \cite{Dreiner} 
and bounds coming from rare meson decays such as the decays of pseudo-scalar and vector mesons
into light neutralino should also be investigated \cite{Dreiner:2009er} in this context. The spectra of
low lying mass eigenstates for the large $f$ case will be shown later for a few benchmark points. 

\item For small $f\sim \mathcal O(10^{-4})$, ${\tilde \chi}^0_7$
is a sterile neutrino state, which is a plausible warm dark matter candidate 
with appropriate relic density. Its mass can be approximated from the 
$8\times 8$ neutralino mass matrix as follows:
\bea
M_N^R &\approx& M_1 \frac{2 f^2 \tan^2\beta}{g^{\prime 2}}.
\label{sterile-mass}
\eea
For a wide range of parameters, the active-sterile mixing angle, 
denoted as $\theta_{14}$, can be estimated as
\bea
\theta_{14}^2 &=& \frac{(m_\nu)_{Tree}}{M_N^R}.
\label{sterile-mixing}
\eea
Furthermore, the sterile neutrino can  be identified with a warm dark matter 
candidate only if the following requirements are fulfilled. These are: (i) it 
should be heavier than 0.4 keV, which is the bound obtained from a model 
independent analysis~\cite{Boyarsky} and (ii) the active-sterile mixing needs 
to be small enough to satisfy the stringent constraint coming from different 
X-ray experiments ~\cite{Boyarsky:2009ix}. 

Under the circumstances, the lightest neutralino-like state is the 
next-to-next-to-lightest eigenstate (${\tilde \chi}^0_6$) of the neutralino 
mass matrix. Its composition is mainly controlled by the parameter $\mu_u$, 
chosen to be rather close to the electroweak scale ($M^D_1, ~M^D_2 > \mu_u$). 
The masses of the lighter neutralino states for this case (small $f$) will be 
presented later.
\end{itemize}

\subsection{The chargino sector}
\label{sec:sub-sec-chargino}
We shall now discuss the chargino sector in some detail as it plays
a crucial role in the decay $h\rightarrow \gamma\gamma$. The relevant Lagrangian
after R-breaking in the AMRB scenario obtains the following form:
\bea
\mathcal L_{ch} &=& M_2 \widetilde w^+ \widetilde w^- + M_2^D \widetilde T_u^+ 
\widetilde w^-+\sqrt 2 \lambda_T v_u \widetilde T_u^+ \widetilde R_d^- 
+g v_u \widetilde H_u^+\widetilde w^- -\mu_u \widetilde H_u^+ \widetilde R_d^-
+\lambda_T v_T \widetilde H_u^+ \widetilde R_d^- \nonumber \\
&-& \lambda_S v_S \widetilde H_u^+ \widetilde R_d^- +g v_a \widetilde w^+ e_L^-
+ M_2^D \widetilde T_u^+ \widetilde w^- + m_e e_R^c e_L^- + h.c.
\eea
The chargino mass matrix, in the basis $(\widetilde w^+,\widetilde T_u^+, \widetilde
H_u^+,e_R^c)$ and $(\widetilde w^-,\widetilde T_d^-,\widetilde R_d^-,e_L^-)$, is 
written as
\bea
M_{c}=\left(
\begin{array}{cccc}
M_{2} & M_{2}^{D} & 0 & gv_a \\
M_{2}^{D} & 0 & \sqrt 2 v_u \lambda_T & 0 \\
g v_u & 0 & -\mu_u-\lambda_S v_S+\lambda_T v_T & 0\\
0 & 0 & 0 & m_e \\
\end{array} \right).
\label{chargino}
\eea
This matrix can be diagonalized by a biunitary transformation, $U M_c V^T=M_D^\pm$. 
The chargino mass eigenstates are related to the gauge eigenstates by these two 
matrices $U$ and $V$. The chargino mass eigenstates (two-component) are written 
in a compact form as
\bea
\chi_i^-&=&U_{ij} \psi_j^-, \nonumber \\
\chi_i^+&=&V_{ij} \psi_j^+,
\label{chargino-mass-two}
\eea
where
\bea
\psi^+_i =  \begin{pmatrix}
\widetilde w^+ \\
\widetilde T_u^+ \\
\widetilde H_u^+ \\
e_R^c
\end{pmatrix},
\hskip 0.2cm
\psi^-_i =  \begin{pmatrix}
\widetilde w^- \\
\widetilde T_d^- \\
\widetilde R_d^- \\
e_L^-
\end{pmatrix}.
\eea
The four-component Dirac spinors can be written in terms of these two-component 
spinors as
\bea
{\widetilde \chi}_i^+ = \left(
\begin{array}{c}
\chi_i^+ \\
\overline \chi^-_i 
\end{array}\right), ~~~~~~~~~~~~~~~~~~(i = 1,...,4).
\eea
It is to be noted that ${\widetilde \chi}_i^c \equiv ({\widetilde \chi}_i^+)^c 
= {\widetilde \chi}_i^-$ is a negatively charged chargino. Hence, the lightest 
chargino (${\widetilde \chi}_4^-$) corresponds to the electron and the structure 
of the chargino mass matrix ensures (see eq.~\ref{chargino}) that the lightest 
mass eigenvalue remains unaltered from the input mass parameter for the electron
, i.e., $m_e = 0.5$ MeV. 

Let us now analyze the composition of different chargino states and how they 
affect the decay width $\Gamma(h\rightarrow \gamma\gamma)$ in this model. Due 
to constraints from the electroweak precision measurements one must consider 
a heavy Dirac wino mass \cite{Kumar-1}. Furthermore, a small tree level 
Majorana neutrino mass demands a mass-degeneracy of the electroweak 
Dirac gauginos as is obvious from eq.~(\ref{neutrino_majorana}). In addition, 
we assume an order one $\lambda_T$ which we use throughout this work for 
numerical purposes. With these,  we observe the following features of the 
next-to-lightest physical chargino state which could potentially contribute 
to $\mu_{\gamma \gamma}$:
\begin{itemize}
\item
In the limit when $M_2^D >> \mu_u$, the next-to-lightest chargino, ${\chi}_3^-$ 
(which is actually the lightest chargino-like state in the MSSM sense), comprises 
mainly of $\widetilde R_d^-$ with a very little admixture of $\widetilde 
T_d^-$ while $\chi_3^+$ is dominated by $\widetilde H_u^+$ with a small admixture 
of $\widetilde w^+$.
\item
For $M_2^D \ll \mu_u$, ${\chi}_3^-$ is predominantly ${\widetilde w}^-$ while 
${\chi}_3^+$ is composed mainly of ${\widetilde T}_u^+$.
\item
Finally, for $M_2^D \approx \mu_u$, ${\chi}_3^-$ is dominantly $\widetilde 
w^-$ and ${\chi}_3^+$ is mostly made up of ${\widetilde T}_u^+$.
\end{itemize}

Apart from the electron, the mass of the chargino states are 
controlled mainly by the parameters $M_2^D$ and $\mu$. We have varied
the input parameters in such a way that the lightest chargino-like state is always 
heavier than 104 GeV \cite{PDG}. The chargino mass spectra corresponding to different 
benchmark points will be presented later.

\section{Contributions to $\mu_{\gamma\gamma}$}
The resonant production of the Higgs boson at the LHC, with the dominant 
contribution coming from gluon fusion, is related to its decay 
to gluons by $\hat \sigma ({gg\rightarrow h)}=\pi^2\Gamma 
(h\rightarrow gg)/8 m_h^3$. Thus, $\mu_{\gamma\gamma}$ can be expressed entirely in 
terms of various decay widths of the Higgs boson as follows 
\cite{Moretti, Moretti-1}:  
\begin{eqnarray}
\mu_{\gamma\gamma}&=&\frac{\sigma (pp\rightarrow h\rightarrow \gamma\gamma)}
{\sigma(pp\rightarrow h\rightarrow \gamma\gamma)^{\rm SM}}, \nonumber \\
&=& \frac{\Gamma (h\rightarrow gg )}{\Gamma (h\rightarrow gg)^{\rm SM}},
\frac{\Gamma^{\rm SM}_{\rm TOT}}{\Gamma_{\rm TOT}}.\frac{\Gamma (h\rightarrow 
\gamma \gamma)}{\Gamma (h\rightarrow \gamma\gamma)^{\rm SM}}.\nonumber \\
&=& k_{gg}. k_{\rm TOT}^{-1}. k_{\gamma \gamma},
\label{eq:mu_gamma_gamma}
\end{eqnarray}
where we use $k_{gg} \equiv \frac{\hat 
\sigma(gg\rightarrow h)}{\hat \sigma(gg\rightarrow h)^{\rm SM}}=
\frac{\Gamma(h\rightarrow gg)}{\Gamma(h\rightarrow gg)^{\rm SM}}$
and $k_{\rm TOT} = \frac{\Gamma_{\rm TOT}}{\Gamma^{\rm SM}_{\rm TOT}}$, $\Gamma_{\rm TOT}$ being
the total decay width of the Higgs boson in the present scenario. The decay of 
$h\rightarrow \gamma\gamma$ is mediated mainly by the top quark and the 
$W^\pm$-loops in the SM and in addition, by top squark, charged Higgs and chargino loops in 
our scenario. In the subsequent discussion we investigate these widths in 
some detail.

As discussed before, note that in this model we have 
integrated out the down type Higgs $(\widehat H_d)$ 
superfield and the sneutrino ${\widetilde \nu}_a$ ($a = 1(e)$) plays the role of 
the down type Higgs boson acquiring a large non-zero $vev$. The sneutrino 
(${\widetilde \nu}_a$) couples to charged leptons (second and third generation) 
and down type quarks via $R$-parity violating couplings which are identified 
with the standard Yukawa couplings. Thus, the couplings of the Higgs boson 
to charged leptons and quarks remain the same as in the MSSM.
This is apparent from the first term given in eq.~(\ref{W-diag}). 

\subsection{The decay $h\rightarrow gg$}
\label{hgg}
The partial width of the Higgs boson decaying to a pair of gluons 
via loops involving quarks and squarks is given by 
\bea
\Gamma(h\rightarrow gg)&=& \frac{G_F \alpha_s^2 m_h^3}
{36\sqrt 2\pi^3} \Big|\sum_{Q} g^h_Q A^h_Q(\tau_q) +\sum_{\widetilde Q} 
g^h_{\widetilde Q} A^h_{\widetilde Q}(\tau_
{\widetilde Q})\Big|^2,
\eea
where $\tau_i=m_h^2/4 m_i^2$, $G_F$ is the Fermi constant, $\alpha_s$ is
the strong coupling constant and 
\bea
A_Q^h(\tau)&=& \frac{3}{2}\left[\tau+(\tau-1)f(\tau)\right]/\tau^2, \nonumber \\
A_{\widetilde Q}^h (\tau)&=& -\frac{3}{4}\left[\tau-f(\tau)\right]/\tau^2,
\eea 
with $f(\tau)$ given by
\bea
f(\tau) & = & \left\{ \begin{array}{ll}
\displaystyle \arcsin^2 \sqrt{\tau} & \tau \le 1, \\
\displaystyle - \frac{1}{4} \left[ \log \frac{1+\sqrt{1-\tau^{-1}}}
{1-\sqrt{1-\tau^{-1}}} - i\pi \right]^2 & \tau > 1.
\end{array} \right.
\label{eq:ftau}
\eea
The couplings are given by
\bea
g^h_Q(u) &=& \frac{\cos\alpha}{\sin\beta}, \nonumber \\
g^h_Q(d) &=& -\frac{\sin\alpha}{\cos\beta}, \nonumber \\
g_{\widetilde Q}^h &=& \frac{m_f^2}{m_{\widetilde Q}^2}g_Q^h \mp \frac{m_Z^2}
{m_{\widetilde Q}^2}(I_3^f-e_f \sin^2 \theta_W)\sin (\alpha+\beta),
\label{eq:top-stop-coupling}
\eea

where the angle $\alpha$ is defined in eq.~(\ref{mixing-angle-alpha}) 
and $\tan\beta=v_u/v_a$.
The couplings of the Higgs boson with the left- and the right-handed squarks are 
exactly the same as in the MSSM. However, one can neglect the mixing between
the left- and the right-handed squarks due to the absence of the $\mu$-term
and the $A$-terms\footnote{Actually, tiny `$A$'-terms are generated because of
the breaking of $R$-symmetry but we can neglect them in the present context.}.  
 
As far as the production of the Higgs boson is 
concerned, we shall show later that a rather light top squark 
with mass around $200-300$ GeV enhances the value of $k_{gg}$ compared to 
the SM. The SM and the MSSM results for the decay $h\rightarrow gg$ 
can be found in~\cite{Djouadi-1,Djouadi-2,Spira}.

\subsection{The decay $h\rightarrow \gamma \gamma$}
\label{hgammagamma}
In the SM, the primary contribution to the decay $h\rightarrow \gamma \gamma$
comes from the $W$ boson loop and the top quark loop with the former 
playing the dominant role. In supersymmetric models, the charged Higgs $(H^{\pm})$, 
top squark $(\widetilde t)$ and the chargino $(\widetilde\chi^{\pm})$ provide extra 
contributions in addition to the $W$ boson and the top quark loop. The authors 
of ref.~\cite{Moretti-1} have noted that the relative strengths of the loop 
contributions involving the vector bosons, the fermions and the scalars with mass around 100 
GeV follow a rough ratio of $8:1.5:0.4$. Nonetheless, a light charged 
Higgs boson ($H^\pm$) could contribute substantially if one considers 
a large $hH^+ H^-$ coupling. However, since the triplet $vev$ is 
small, the contribution of the triplet to the charged Higgs state is negligible.  
On the other hand, charginos in loop could enhance the $h\rightarrow 
\gamma\gamma$ decay width, in particular, when they are light and/or  
diagrams involving them interfere constructively with the $W$-mediated loop 
diagram.

The Higgs to diphoton decay rate can be written down as  \cite{Djouadi-1}
\bea
\Gamma(h\rightarrow \gamma\gamma) &=& \frac{G_F \alpha^2 m_h^3}
{128\sqrt 2 \pi^3}\Big|\sum_f N_c Q_f^2 g_f^h A_{1/2}^h+g_{hW^+W^-} A_{1}^h+
g_{hH^+H^-} A_{0}^{h}+\sum_{\widetilde c} g_{h \widetilde\chi_i^+ \widetilde\chi_j^-} 
A_{1/2}^h \nonumber \\
&+& \sum_{\widetilde f} N_c e_{\widetilde f}^2 g_{h\widetilde f\widetilde f}
A_{0}^h\Big|^2,
\eea
where
\bea
A_1^h &=& -[2\tau^2+3\tau+3(2\tau-1)f(\tau)]/\tau^2, \nonumber \\
A_{1/2}^h &=& 2[\tau+(\tau-1)f(\tau)]/\tau^2, \nonumber \\
A_0^h &=& -[\tau-f(\tau)]/\tau^2,
\eea
with the loop functions already defined in eq.~(\ref{eq:ftau}). The relevant 
couplings are given by,
\bea
g_{h\overline u u}&=& \frac{\cos\alpha}{\sin\beta}, \nonumber \\
g_{h\overline d d}&=& -\frac{\sin\alpha}{\cos\beta}, \nonumber \\
g_{hWW} &=& \sin(\beta-\alpha), \nonumber \\
g_{hH^+H^-}&=& \frac{m_W^2}{m_{H^{\pm}}^2}[\sin(\beta-\alpha)+
\frac{\cos2\beta \sin(\beta+\alpha)}{2\cos^2\theta_W}],\nonumber \\
g_{h\widetilde f\widetilde f}&=& \frac{m_f^2}{m_{\widetilde f}^2}g_{hff}\mp
\frac{m_Z^2}{m_{\widetilde f}^2}[I_3^f-e_f \sin^2\theta_w]
\sin(\alpha+\beta), \nonumber \\
g_{h\widetilde \chi^+_i \widetilde \chi^-_j}&=& 2 \frac{m_W}{m_{\widetilde c_k}}
(\xi_{ij}\sin\alpha-\eta_{ij}\cos\alpha).
\label{higgs-couplings-gam-gam}
\eea
Here $\xi_{ij}=-\frac{1}{\sqrt 2}V_{i1}U_{j4}$ and 
$\eta_{ij}=-\frac{1}{\sqrt 2}\left(\frac{\sqrt 2\lambda_T}{g} U_{i3}
V_{j2}+U_{i1}V_{j3}\right)$. The masses which appear in the denominator
of the couplings given above, represent physical masses propagating in the loop. For example,
$m_{\widetilde c_k}$ are the physical chargino masses, $m_{\widetilde f}$ are the physical masses of the
sfermions and so on.
We present the complete set of Higgs-chargino-chargino interaction vertices 
in Appendix~A\footnote{For the MSSM case see ~refs. 
\cite{Spira,Spira-1}.}.

As noted earlier, the largest contribution in the Higgs decay rate to two photons 
comes from the $W$ boson loop. Similar to the MSSM, the $hWW$ coupling gets 
modified by the factor $\sin(\beta-\alpha)$. Hence, in order to have a significant 
contribution from the $W$ boson loop in our model, the angles $\alpha$ and $\beta$ 
need to be aligned in such a way that one obtains a large value of 
$\sin(\beta-\alpha)$, which can be achieved in the decoupling regime, i.e., the coupling
to the lightest Higgs boson becomes SM like. 

\begin{figure}[htb]
\centering
\includegraphics[height=2.55in,width=2.5in]{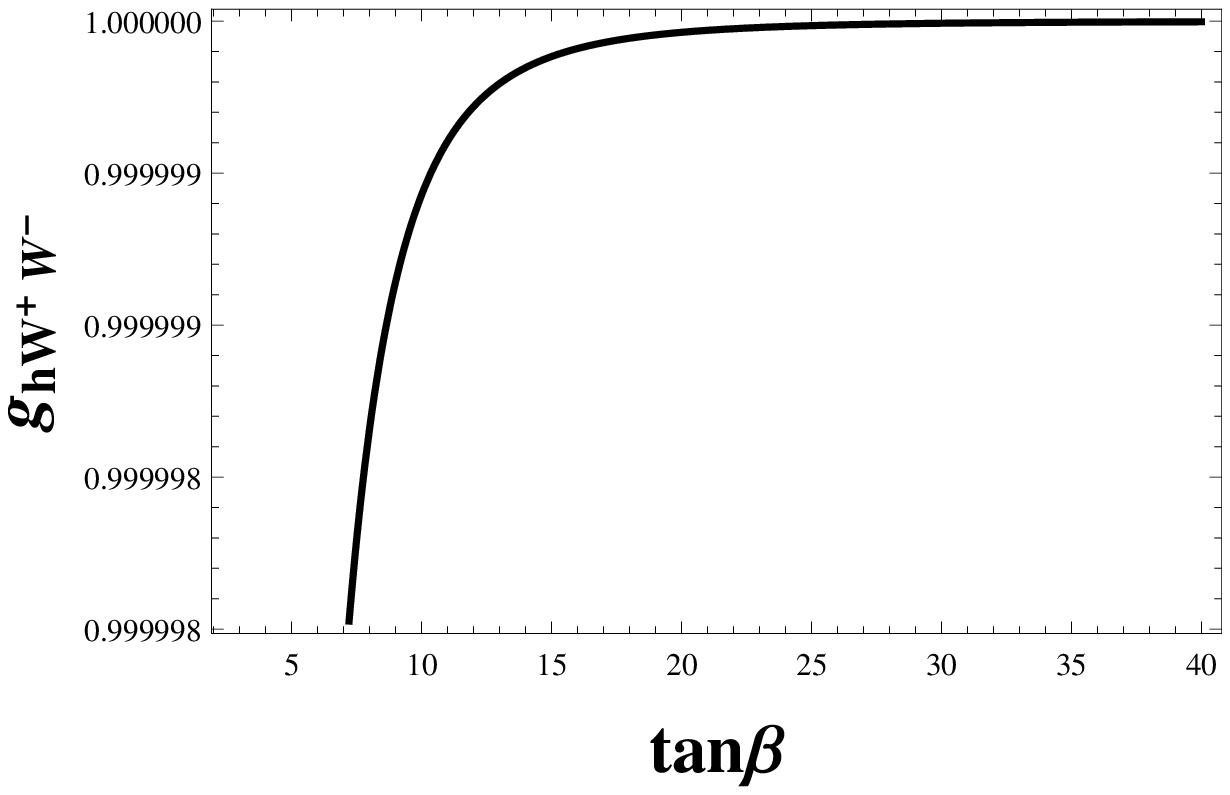} 
\qquad
\includegraphics[height=2.5in,width=2.8in]{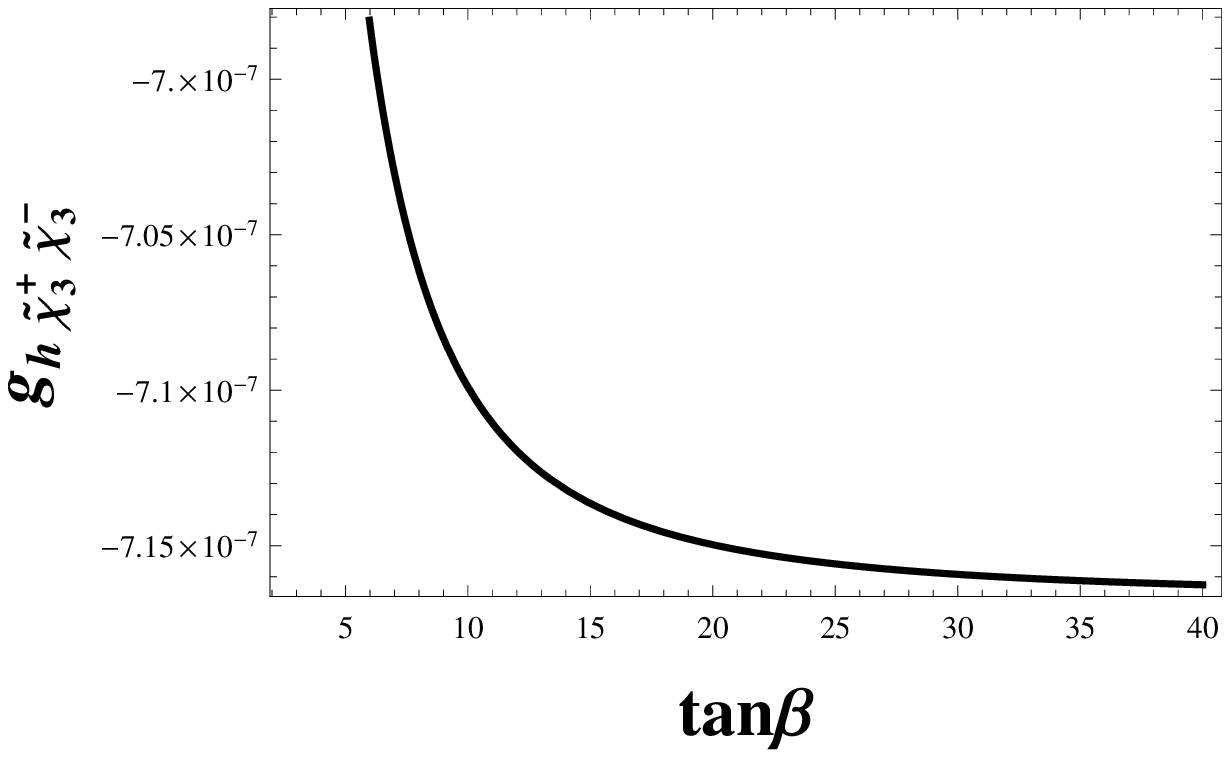}      
%
\caption{Couplings of the lightest Higgs boson to a pair of $W$-bosons 
(left) and to a pair of light charginos 
($\mathbf{\widetilde{\chi}_3^\pm}$) (right).}
\label{fig:higgs-couplings}
\end{figure}

\begin{figure}[htb]
\begin{center}
\includegraphics[height=3in,width=4in]{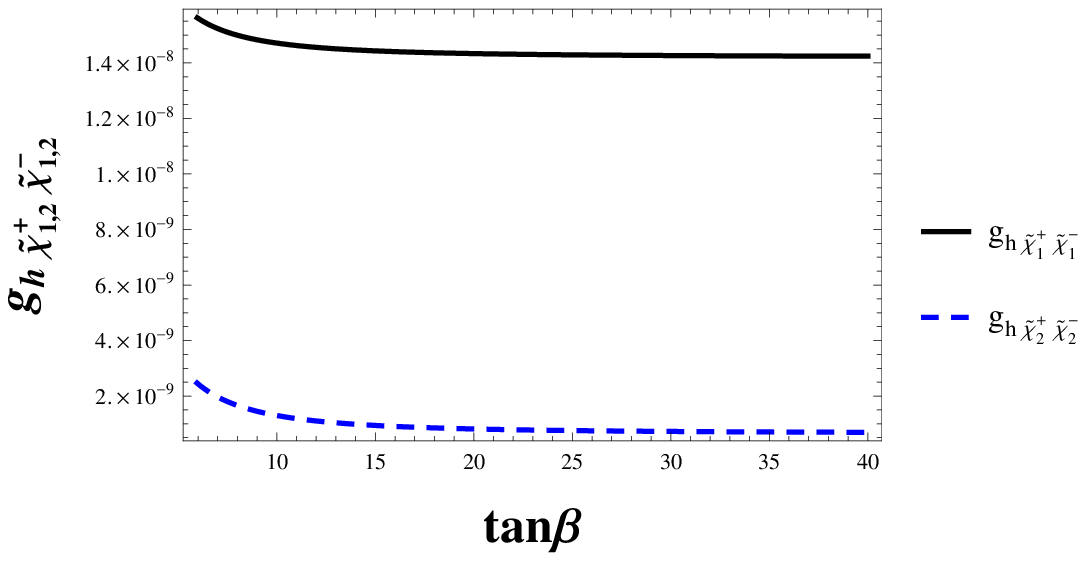}
\caption{\label{higgs_heavy_charginos} Couplings of the Higgs 
boson to heavier charginos. The thick black line represents the coupling 
to the heaviest chargino {\bf ($\mathbf{\widetilde{\chi}_1^\pm}$)} whereas 
the blue dashed one represents the same to the chargino immediately 
lighter to it {\bf ($\mathbf{\widetilde{\chi}_2^\pm}$)} .}
\end{center}
\label{fig:higgs_heavy_charginos}
\end{figure}

\begin{figure}[htb]
\begin{center}
\includegraphics[height=3.0in,width=4.0in]{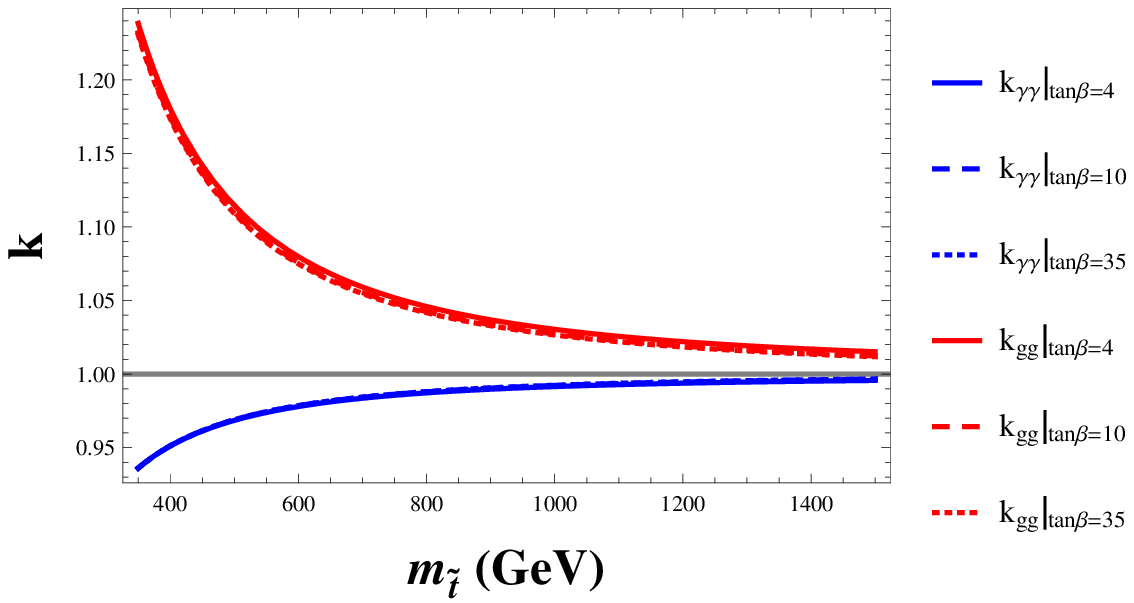} 
\caption{\label{fig:k_stop} Variations of $k_{gg}$ (in red) and 
$k_{\gamma\gamma}$ (in blue) as functions of $M_{\widetilde t}$
for $\tan\beta=4$, 10, 35.}
\end{center}
\end{figure}

In fig.~\ref{fig:higgs-couplings}, we illustrate the variations of 
the couplings $g_{hW^+ W^-}$ and $g_{h{\widetilde \chi}^+_3 {\widetilde \chi}^-_3}$, 
which might play important roles in the decay $h\rightarrow \gamma\gamma$. 
We choose $M_1^D = 1.5$ TeV, $\mu_u = 200$ GeV, $m_{3/2}=10$ GeV, $m_{\widetilde t} 
= 500$ GeV, $v_S = 10^{-4}$ GeV, $v_{T} = 10^{-3}$ GeV and retain a near 
degeneracy between the Dirac gaugino masses with $\epsilon \equiv (M_2^D - M_1^D) 
= 10^{-1}$ GeV, with $f = 0.8$ and $B\mu_L=-(400)^2 (\text{GeV})^2$. From the 
left panel of fig. \ref{fig:higgs-couplings} 
we observe that the $hWW$ coupling is almost SM like as we are essentially working
in the decoupling limit. This implies that the $W$-loop contribution in the
$h\rightarrow\gamma\gamma$ process remains almost unchanged with varying $\tan\beta$.
%
%
On the other hand, as $\mu_u << M^D_{1,2}$, the next-to-lightest chargino state 
is dominantly controlled by the $\mu_u$ parameter. For this case, the coupling 
$g_{h{\widetilde \chi}^+_3 {\widetilde \chi}^-_3}$ is plotted as a function of 
$\tan\beta$ in the right panel of fig. \ref{fig:higgs-couplings}. One can 
clearly see that $g_{h{\widetilde \chi}^+_3 {\widetilde \chi}^-_3}$ is already 
much suppressed compared to $g_{hW^+ W^-}$, for the entire range of $\tan\beta$. 
From the expression for $g_{h{\widetilde \chi}^+_3 {\widetilde \chi}^-_3}$ in 
eq.~(\ref{higgs-couplings-gam-gam}) it is straightforward to verify that this 
coupling remains very much suppressed for all the different cases mentioned in 
section \ref{sec:sub-sec-chargino}. The Higgs boson couplings to heavier 
charginos are also highly suppressed as can be seen from fig. 
\ref{higgs_heavy_charginos}. Thus, the contribution of charginos in 
$\Gamma(h \rightarrow \gamma \gamma)$ would, in any case, be insignificant. 
Referring back to eq.~(\ref{eq:mu_gamma_gamma}), we are now in a position 
to have some quantitative estimates of the quantities $k_{gg}$ and 
$k_{\gamma\gamma}$ which control the signal strength $\mu_{\gamma\gamma}$. In 
fig. \ref{fig:k_stop} we illustrate their variations ($k_{gg}$ in red and 
$k_{\gamma\gamma}$ in blue) as functions of the mass of the top squark for various
values of $\tan\beta$. We observe that $k_{gg}$ is not at all sensitive 
to $\tan\beta$ (all three curves in red for three $\tan\beta$ values are 
found to be overlapping).
This is since we considered $gg\rightarrow h$ production via loops
involving the top quark and the top squark. 
The couplings involved there carry a 
factor $\cos\alpha/\sin\beta$, which varies only 
marginally with respect to $\tan\beta$. 
Similarly $k_{\gamma\gamma}$ also remains insensitive with $\tan\beta$.
The reason being, $\Gamma (h\rightarrow\gamma\gamma)$ receives major contribution
from the $W$ boson induced loop where the involved coupling goes as $\sin(\beta-\alpha)$.
As shown vividly in the left panel of fig. \ref{fig:higgs-couplings} that 
$hWW$ coupling remains almost unchanged with $\tan\beta$. As a result, $k_{\gamma
\gamma}$ shares the same feature as $k_{gg}$ as far as variation with respect to $\tan\beta$ is
concerned.

It is observed that for for light top squarks, $k_{gg}$ gets enhanced by a 
considerable amount. However, in that very region , $k_{\gamma\gamma}$ is rather 
small for small $\tan\beta$, and  it becomes somewhat larger for higher 
$\tan\beta$. However, it is found that $k_{gg} > 1$ while $k_{\gamma\gamma} < 1$, 
all through. We have also checked that the illustrated variations of 
$k_{gg}$ and $k_{\gamma\gamma}$ are following their respective gross
trends in the MSSM closely in the limit of zero left-right 
mixing in the scalar sector.
Note that for this plot we have not incorporated the constraints 
from the mass of the Higgs boson and the requirement of having no tachyonic 
scalar states. In section \ref{sec:impact-numerical}, while discussing the quantitative impact 
of the recent LHC results on such a scenario,
we present results of detailed scan of the parameter 
space by including all these constraints. 
%

All the previous plots consider a large values of `$f$' 
($f$ $\sim \mathcal O(1)$) for which one obtains a large tree level 
correction to the Higgs boson mass as well as an appropriate 
mass for the active neutrino at the tree level. We adopt such a
scenario with relatively large values of `$f$' in our study of the
Higgs boson decay rates which we present in the next subsection.
%
\subsection{Higgs boson decaying to charginos and neutralinos}
In the presence of much lighter charginos and neutralinos (as discussed in 
sections \ref{sec:neutralino-r-breaking} and \ref{sec:sub-sec-chargino}),
an SM-like Higgs boson with mass around 125 GeV could undergo decays to a pair 
of these states. We study these things in detail in this section.

It has been noted in section \ref{sec:neutralino-r-breaking}, that the smallest 
eigenvalue ($m_{{\widetilde \chi}^0_8}$) of the neutralino mass matrix corresponds 
to the neutrino mass. The next-to-lightest neutralino (${\widetilde \chi}^0_7$) turns 
out to be a bino-like neutralino (the sterile neutrino) for large (small) values of 
`$f$'. Moreover, the mass of the next-to-next-to-lightest neutralino state 
(${\widetilde \chi}^0_6$) is mostly controlled by $\mu_u$. Since we have chosen 
$\mu_u$ to be very close to the electroweak scale, the Higgs boson decay to a pair 
of ${\widetilde \chi}^0_6$ is not possible. 
The presence of light neutralino states may enhance the invisible decay
width of the Higgs boson considerably. Amongst them, the most dominant
contribution comes from $h$-$\widetilde\chi_7^0$-$\widetilde\chi_8^0$ coupling.
However, a detailed study reveals that the contribution of this coupling
is not substantial and hence the corresponding decay width is rather small.
\begin{figure}[htb]
\begin{center}
\includegraphics[height=3.3in,width=3.8in]{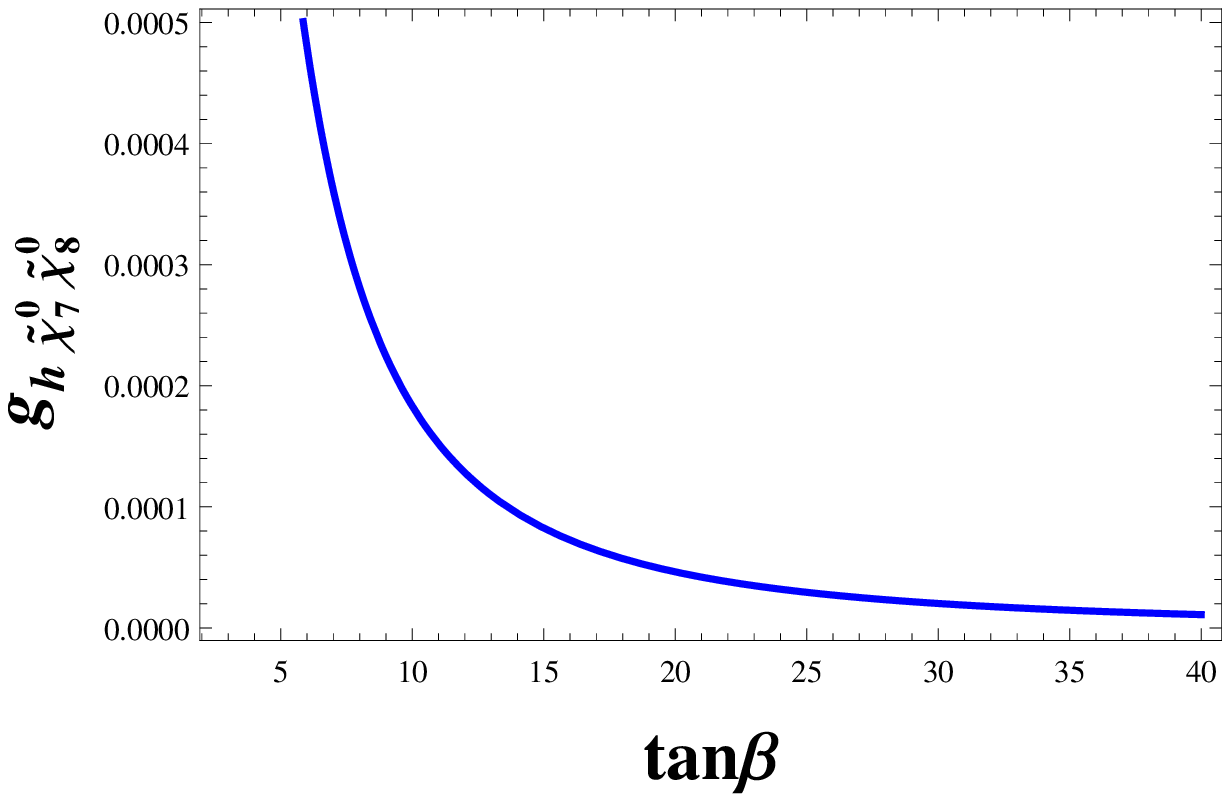}
\caption{\label{higgs-decay-neutralino}
Variation of the $h-\widetilde\chi_7^0-\widetilde\chi_8^0$ coupling
as a function of $\tan\beta$.
} 
\end{center}
\end{figure}
It is clear from fig.~\ref{higgs-decay-neutralino} 
that the $h$-$\widetilde\chi_7 ^0$-$\widetilde \chi_8^0$ coupling grows for
small $\tan\beta$. This is essentially because for smaller values
of $\tan\beta$, the sneutrino component of the lightest Higgs boson mass eigenstate
is large, which results in a slightly larger value of this coupling.
This fact also shows up for the invisible decay widths of the Higgs boson,
which we will discuss later.

On the other hand, the lightest chargino eigenstate (${\widetilde \chi}^\pm_4$)
corresponds to the electron. The mass of the next-to-lightest chargino 
(${\widetilde \chi}^\pm_3$) is again controlled by $\mu_u$ if $\mu_u < M_2^D$. 
Thus, decay of the Higgs boson to a pair of ${\widetilde \chi}^\pm_3$ is not possible.
The most general expressions for the partial widths of the Higgs boson 
decaying to a pair of neutralinos ($\Gamma(h \rightarrow \widetilde\chi_i^0\widetilde
\chi_j^0$)) or a pair of charginos ($\Gamma ({h\rightarrow \widetilde\chi_i^{+}
\widetilde\chi_j^{-}}$)) can be found in the Appendix~A and B. 

\subsection{The total decay width of the Higgs boson}
\label{htot}
In this section we collect the partial decay widths of the lightest Higgs boson
that dominantly contribute to its total decay width. The latter is thus given
by\footnote{We neglect the rare decay modes like $H \to Z \gamma$, $\gamma^* \gamma$, 
$\mu^+\mu^-$, $e^+e^-$ etc.}
\bea
\Gamma_{\rm TOT}&=&\Gamma(h\rightarrow b\overline b)+\Gamma(h\rightarrow 
\tau \overline\tau)+\Gamma(h\rightarrow gg)+\Gamma(h\rightarrow WW^*) 
+\Gamma(h\rightarrow ZZ^*) \nonumber \\
&+& \Gamma(h\rightarrow \gamma \gamma) 
+\Gamma (h\rightarrow \widetilde\chi_i^0\widetilde\chi_j^0)
+\Gamma (h\rightarrow \widetilde\chi^+_i\widetilde\chi^-_j).\nonumber \\
\eea
For completeness, we present here the analytical expressions 
for all the decay rates which go into our analysis but 
were not presented earlier. These are as follows:
\bea
\Gamma(h\rightarrow b\overline b)&=& \frac{3 G_F m_b^2 m_h}{4\pi \sqrt 2} 
\Big(\frac{\sin\alpha}{\cos\beta}\Big)^2
\Big[1-\frac{4 m_b^2} {m_h^2}\Big]^{3/2}, \nonumber \\
\Gamma(h\rightarrow \tau\overline\tau)&=& \frac{G_F m^2_\tau m_h}{4\pi \sqrt 2} 
\Big(\frac{\sin\alpha}{\cos\beta}\Big)^2
\Big[1-\frac{4 m_\tau^2} {m_h^2}\Big]^{3/2}, \nonumber \\
\Gamma(h\rightarrow WW^*)&=& \frac{3G_F^2 m_W^4
m_h}{16\pi^3}\sin^2(\alpha-\beta)
R\Big(\frac{m_W^2}{m_h^2}\Big), \nonumber \\
\Gamma(h\rightarrow ZZ^*)&=& \frac{3G_F^2 m_Z^4 m_h} {16\pi^3} 
\Big[\frac{7}{12}-\frac{10}{9}
\sin^2\theta_W+\frac{40}{27}\sin^4\theta_W\Big]
R\Big(\frac{m_Z^2}{m_h^2}\Big).
\eea
The function $R(x)$ is defined as~\cite{Spira,Keung,Rizzo} 
\bea
R(x)&=& 3\frac{(1-8x+20 x^2)}{\sqrt (4x-1) \arccos
\Big(\frac{3x-1}{2x^{3/2}}\Big)}-\Big(\frac{1-x}{2x}\Big)
(2-13x+47x^2)-\frac{3}{2}(1-6x+4x^2) \log x.\nonumber \\
\eea
The Higgs boson decay rates to charginos and neutralinos are shown in
Appendix A and B respectively. The recent CMS analysis constrains the
total decay width of the Higgs boson to be less than 14 MeV or so~\cite{Khachatryan:2014iha}.
In the subsequent sections we present the numerical results of our analysis 
pertaining to the diphoton signal strength $\mu_{\gamma \gamma}$ and subject 
this to important experimental findings.
\section{Impact of the LHC results}
\label{sec:impact-numerical}
%
In this section, we discuss the impact of the findings from the LHC pertaining 
to the Higgs sector on the scenario under discussion. As pointed out earlier, 
two broad scenarios based on the magnitude of `$f$' worth special attention: 
the scenario with large `$f$' ($\sim\mathcal O(1)$) and the one for which
`$f$' is rather small.
%
\subsection{The case of large neutrino Yukawa coupling, $f\sim\mathcal O(1)$}
A large neutrino Yukawa coupling ($f\sim \mathcal O(1)$) already enhances 
the tree level Higgs boson mass. Thus, such a scenario banks less on 
large radiative contributions from the top squark loops to uplift the 
same. 
Further, an appropriately small 
tree level Majorana neutrino mass (the lightest neutralino) can be obtained 
along with a light bino-like neutralino ($\widetilde{\chi}^0_7$, the 
next-to-lightest neutralino) once $R$-symmetry is broken explicitly, via anomaly 
mediation. The mass of this neutralino is essentially controlled by the $R$-symmetry 
breaking Majorana mass term of the $U(1)$ gaugino (the bino), i.e., $M_1$, and 
hence related to the gravitino mass $m_{3/2}$. Since we assume $m_{3/2} 
\sim 10$ GeV, the next-to-lightest neutralino acquires a mass of the order of a 
few hundred MeV. The presence of such a light bino like neutralino implies an additional 
contribution to the total decay width of the Higgs boson. 
We also looked at the diphoton signal strength $\mu_{\gamma\gamma}$ and
compared it with the latest ATLAS and CMS results.
\subsubsection{Invisible branching ratio of the Higgs boson}
\begin{figure}[htb]
\begin{center}
\includegraphics[height=3.3in,width=3.8in]{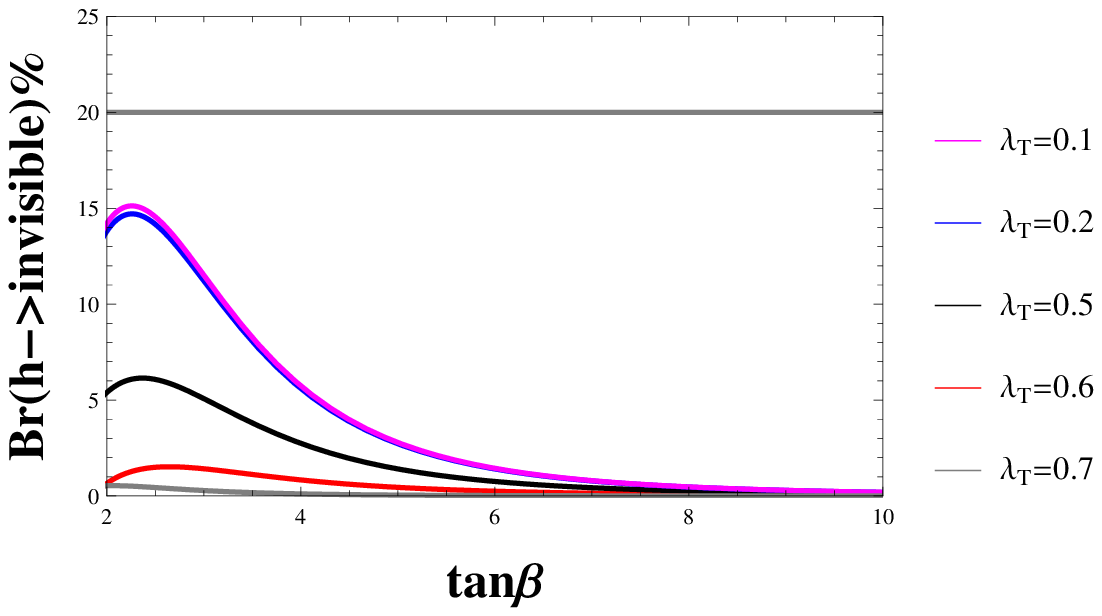}
\caption{\label{invisible_br}
The lightest Higgs boson invisible branching ratio as a function of $\tan\beta$ 
for different values of $\lambda_T$. The horizontal line corresponds to the upper limit
on the invisible branching ratio from model independent analysis \cite{Cheung:2014noa}.} 
\end{center}
\end{figure}
To take into account the constraints coming from the invisible decay branching 
ratio of the lightest Higgs boson
we have fixed $M_1^D = M_2^D = 1.5$ TeV, $\mu_u = 200$ GeV, (i.e., $M_1, M_2 
<< \mu_u$), $m_{3/2} = 10$ GeV, $m_{\widetilde t} = 500$ GeV, $v_S = 10^{-4}$ GeV 
and $v_T = 10^{-3}$ GeV with $f=0.8$, $B\mu_L=-(200)^2 (\text{GeV})^2$. As
discussed earlier, the partial decay
width of the Higgs boson decaying to a neutrino and a neutralino ($h\rightarrow
\widetilde\chi_7^0\widetilde\chi_8^0$) could essentially contribute to the invisible
final state. This can be understood from the fact that although $\widetilde\chi_7^0$
would undergo $R$-parity violating decays $\widetilde\chi_7^0\rightarrow q\overline q
\nu$, $e^+ e^- \nu$, $\nu\nu\nu$, $q\overline q^{\prime} e^-$, where $q$, $q^{\prime}$
are the SM light quark states from the first two generations, these decay modes
involve very small couplings
and as a result, the decay length happens to be much larger than the collider 
dimension. Therefore, the LSP neutralino contributes to missing energy (MET) signals
\cite{Porod:2000hv}. Note that $\Gamma(h\rightarrow\widetilde\chi_7^0\widetilde\chi_7^0)$
is negligibly small because of suppressed $h$-$\widetilde\chi_7^0$-$\widetilde\chi_7^0$ 
coupling for a bino-dominated, $\widetilde\chi_7^0$.

We observe from fig. \ref{invisible_br} that this partial decay width is comparatively
larger for smaller values of $\tan\beta$ and $\lambda_T$. However, it is clear
that the presence of a bino-like neutralino state is not yet constrained
from the invisible decay mode of the Higgs boson in our scenario with all
the curves staying well below the experimentally derived~\cite{Cheung:2014noa} upper bound of
$\sim 20\%$ for the invisible branching fraction of the Higgs boson.
\subsubsection{The signal strength $\mu_{\gamma\gamma}$}
\label{sec:signal-strength}
\begin{figure}[htb]
\begin{center}
\includegraphics[height=3.3in,width=3.8in]{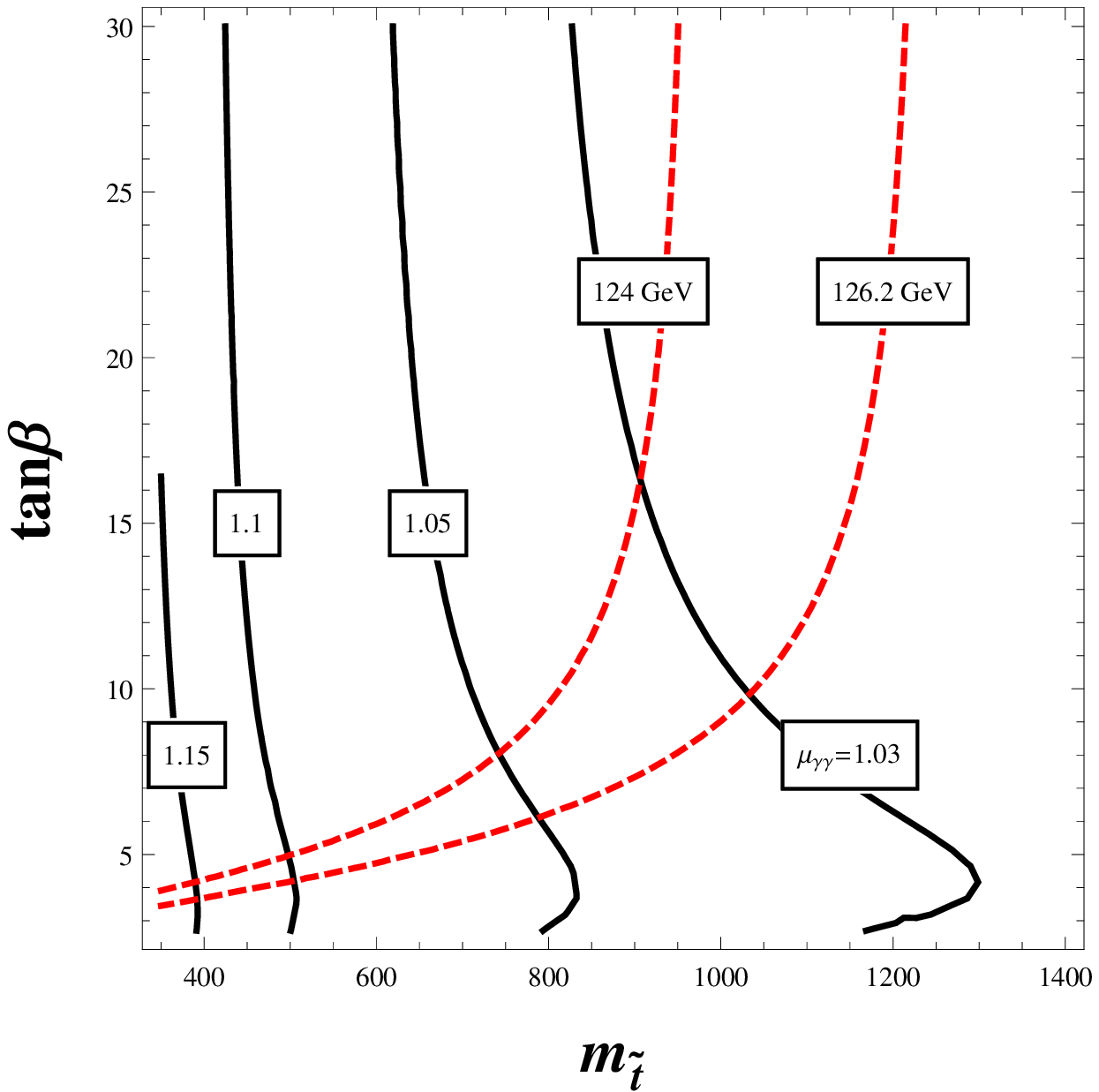}
\caption{\label{fig:mugg_1}
Contours of various fixed values of $m_h$ (124 GeV and 126.2 GeV), 
$\mu_{\gamma\gamma}$ and $k_{TOT}$ 
in the $m_{\widetilde{t}}$ --$\tan\beta$ plane. $\lambda_T$ and $f$ are 
fixed at 0.45 and 0.8, respectively. Other parameters are set at the values 
as mentioned in the text. 
}
\end{center}
\end{figure}

It is now important to analyse the signal strength corresponding to the 
$h\rightarrow \gamma\gamma$ channel. 
In fig. \ref{fig:mugg_1} we fix $\lambda_T = 0.45$, and $f$ = 0.8, with all other 
parameters held at the values mentioned in section~\ref{hgammagamma}.
The red dashed lines represent the contours of $m_h$ = 124 GeV and 126.2 GeV
respectively and enclose the experimentally allowed range of $m_h$. The black thick 
lines are the contours of fixed $\mu_{\gamma \gamma}$ with values 1.15, 1.1, 1.05 and 1.03 
respectively. 
Figure \ref{fig:mugg_1} shows 
that there is an available region of parameter space consistent with the latest experimental 
findings involving $m_h$ and $\mu_{\gamma\gamma}$. 
Relatively low values of the top squark mass results in an increase of the cross
section for the resonant Higgs boson production through gluon fusion and thus enhances $\mu_{\gamma\gamma}$.
On the other hand $\mu_{\gamma\gamma}$ is almost insensitive to $\tan\beta$ for
$\tan\beta\geq 15$. This is because $hb\overline b$ coupling (which controls the total decay
width of the Higgs boson in a significant way) becomes independent of $\tan\beta$
for larger values of this parameter.
\begin{figure}[htb]
\begin{center}
\includegraphics[height=3.3in,width=3.8in]{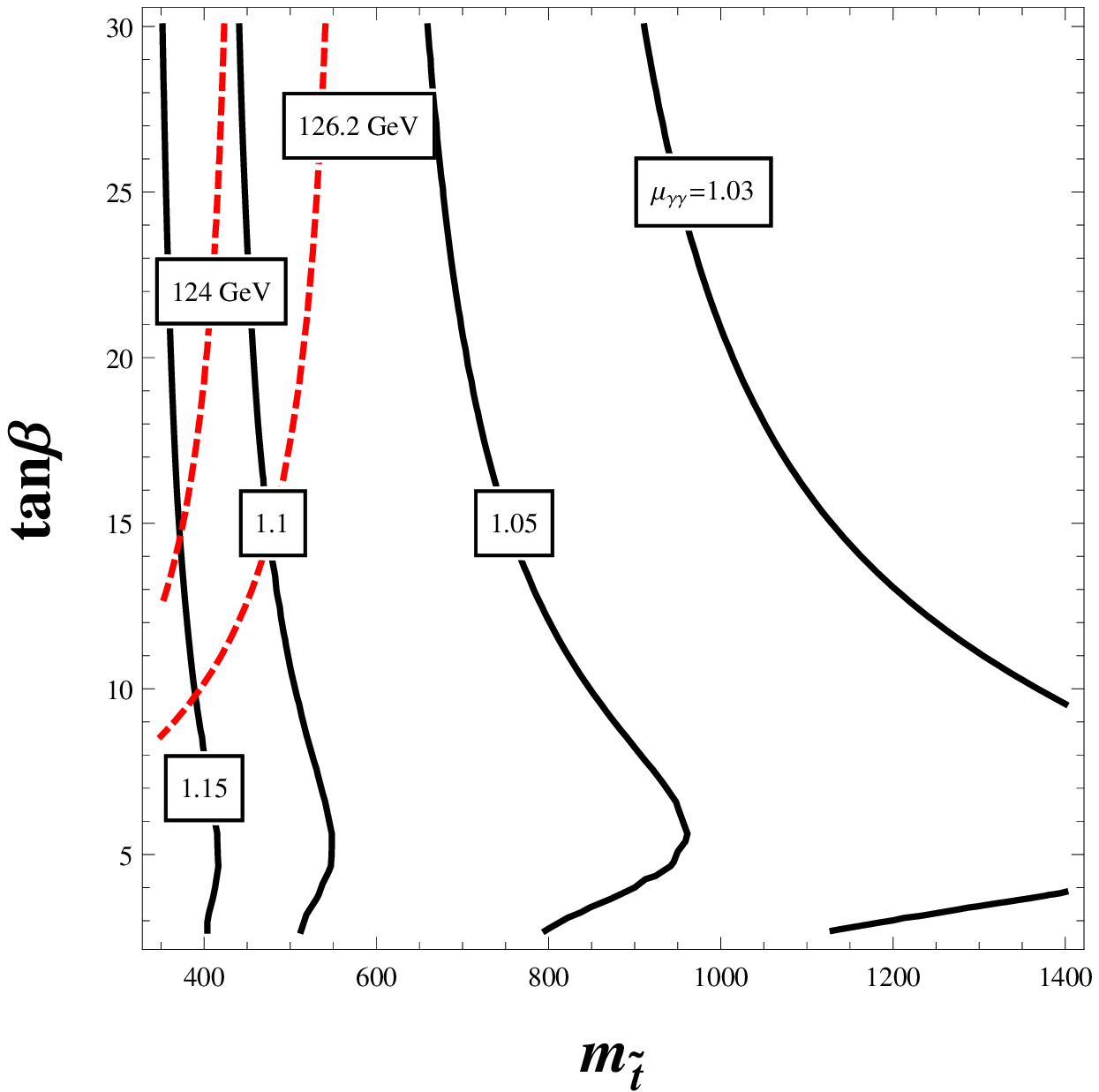}
\caption{\label{fig:mugg_2}
Same as in fig. \ref{fig:mugg_1} but with $\lambda_T=0.5$ and $f=1$.}
\end{center}
\end{figure}

Figure \ref{fig:mugg_2} addresses the same issue but with $f=1$ and $\lambda_T 
= 0.5$. Since a larger value of $\lambda_T$ already provides a 
significant contribution to 
the Higgs boson mass via radiative correction, only light top squarks are 
compatible with the measured range of $m_h$. Moreover, a larger value of `$f$' 
implies a larger $\tan\beta$ to have the Higgs boson mass in the correct range. 
It is pertinent to mention that these plots use spectra of particles which are 
consistent with the lower bound on the lightest chargino mass ($> 104$ GeV, 
from the LEP experiments) and are also free from tachyonic scalar states.

\subsubsection{Relative signal strengths in different final states}
\label{sec:relative-strength-large-f}
\begin{figure}%
    \centering
    {{\includegraphics[width=7cm]
    {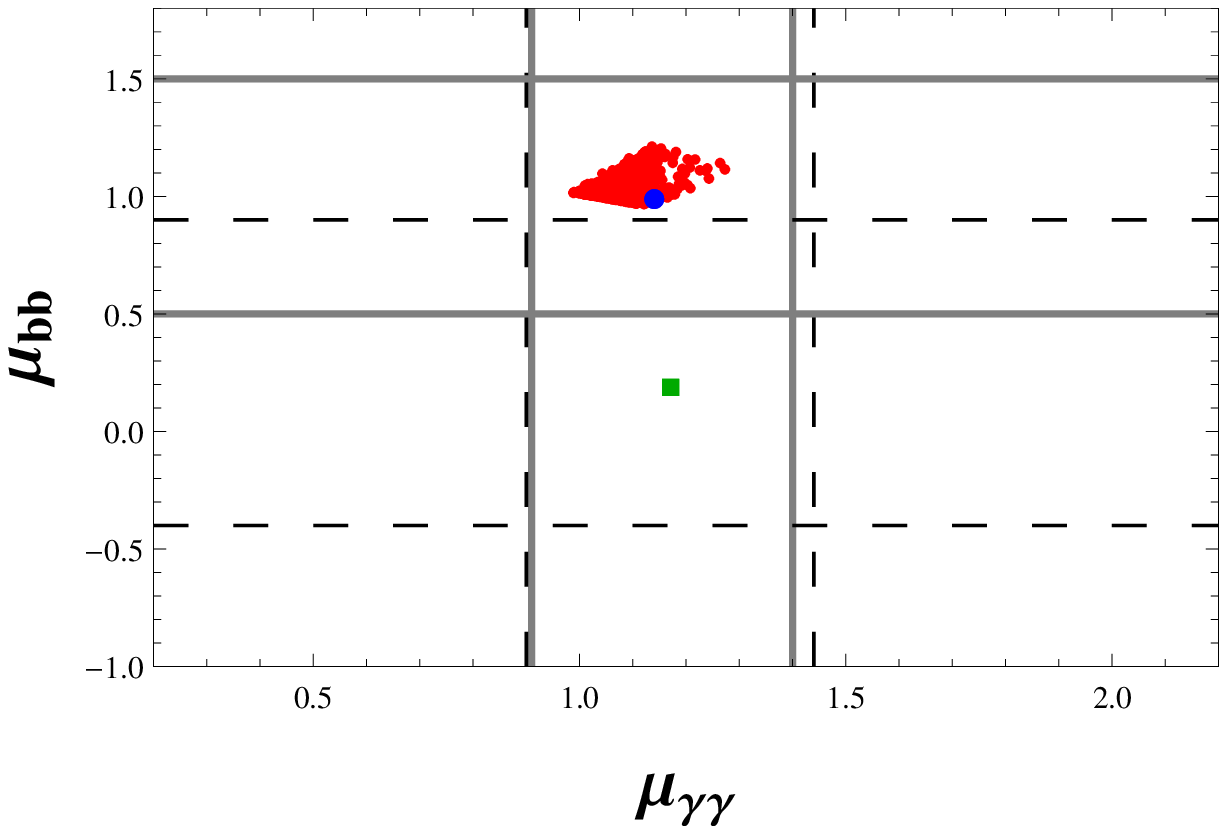} }}%
    \qquad
    {{\includegraphics[width=7cm]
    {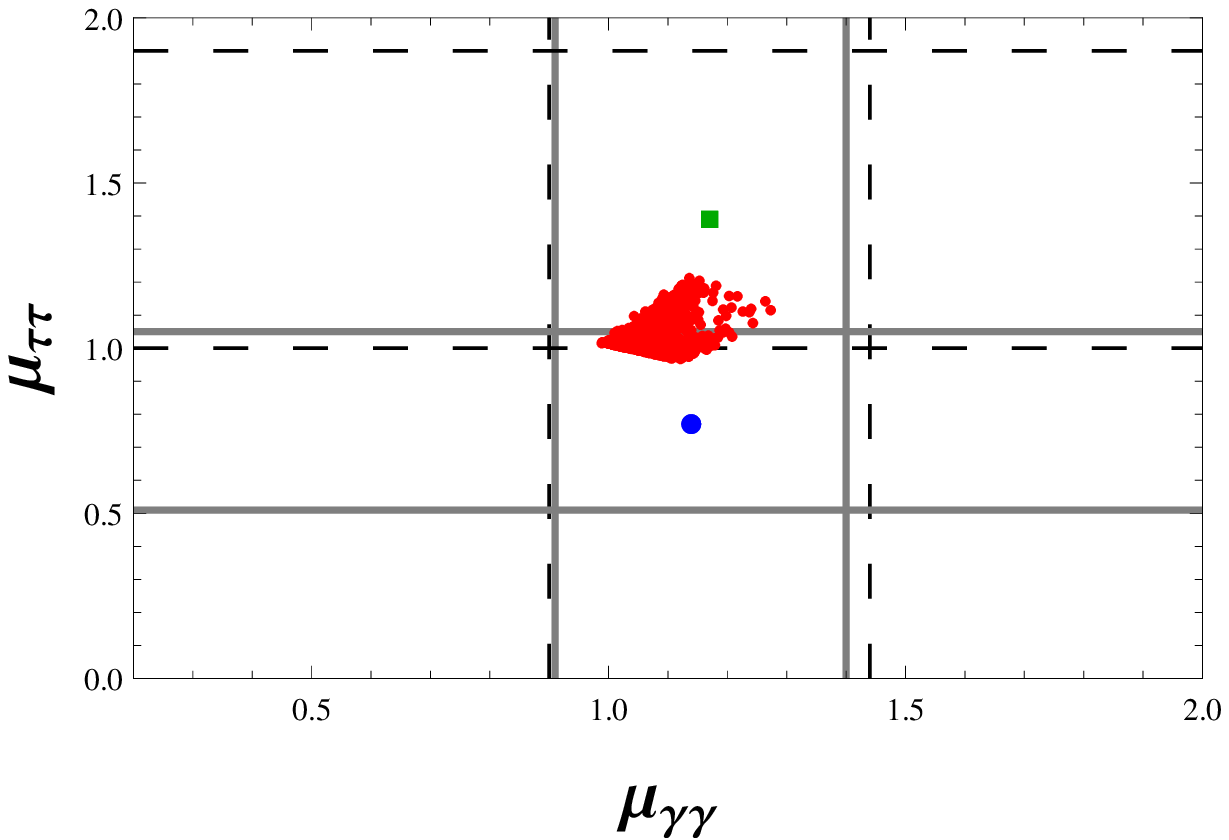} }}%
    \qquad
    {{\includegraphics[width=7cm]
    {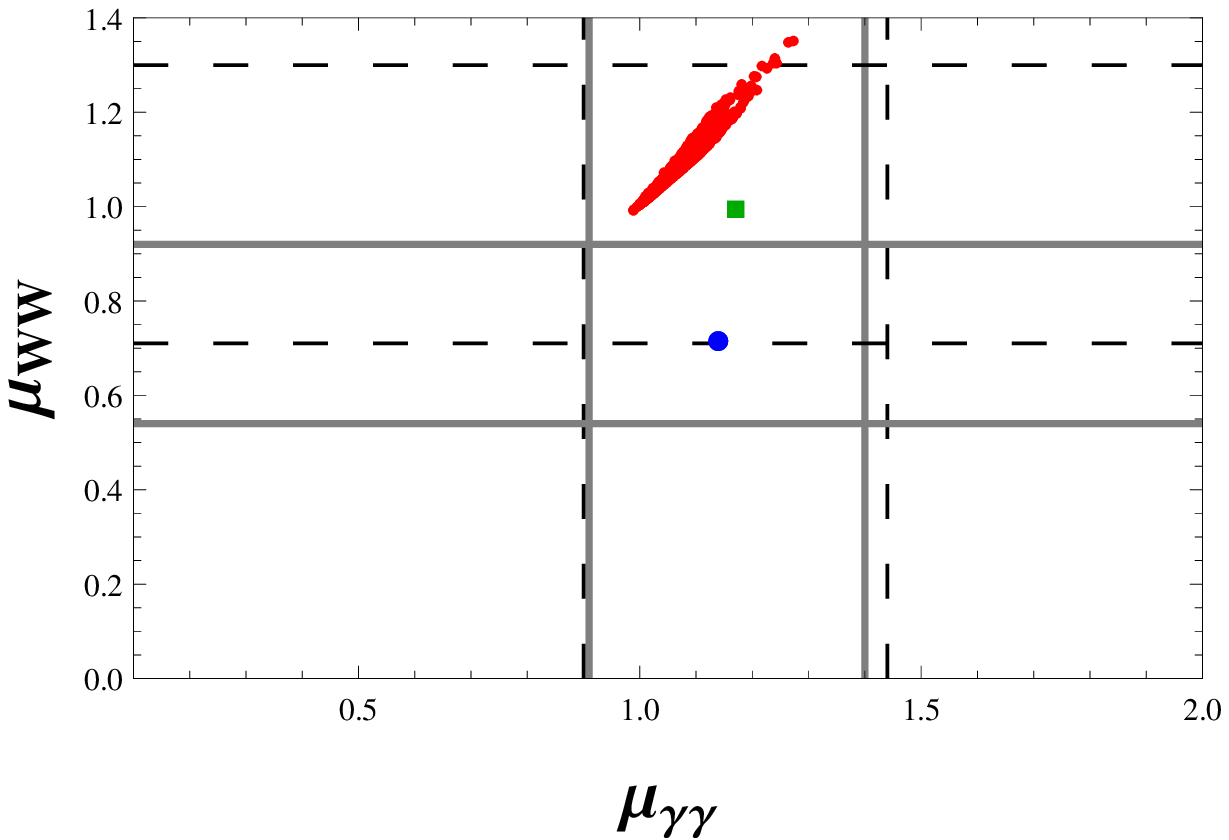} }}%
    \qquad
    {{\includegraphics[width=7cm]
    {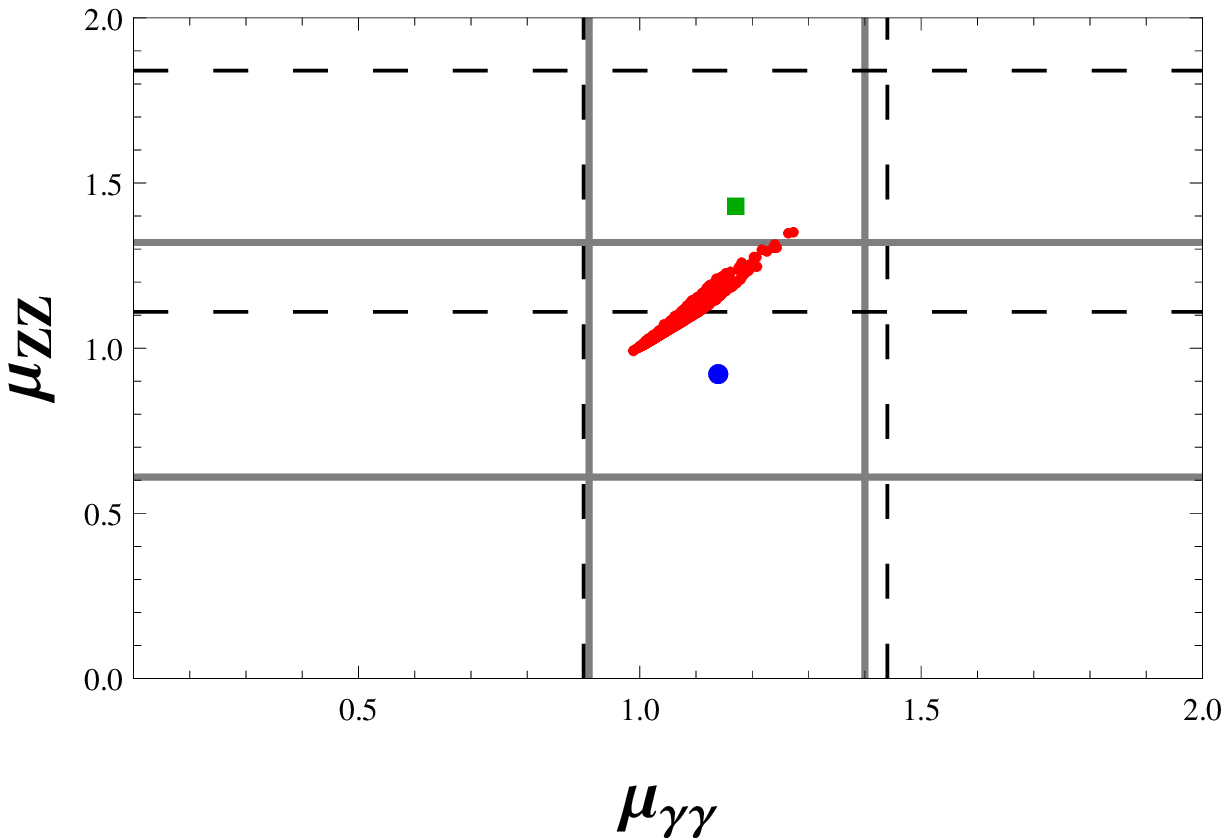} }}%
    \caption{
Bands representing mutual variation of relative signal strengths in various 
possible final states arising from the decay of the lightest Higgs boson as 
obtained by scanning the parameter space of the scenario under consideration. 
The ranges of different parameters used in the scan are as follows: 
$10<\tan\beta<40$, $350~ {\mathrm{GeV}} < m_{\widetilde t} < 1.5~ \rm{TeV}$, 
$0.1 < f < 1$ and $0.1 < \lambda_T < 0.55$. The solid grey lines give 
1-$\sigma$ ranges from the MVA based analysis (main analysis) performed by 
the CMS collaboration (blue circles represent the respective central values) 
whereas the dashed grey lines represent the corresponding results from the 
ATLAS collaboration (green squares represent the respective central values). 
}
\label{fig:comp-signal-strength}%
\end{figure}

In this subsection we briefly discuss how other final states arising from the 
lightest Higgs boson are expected to be affected in our scenario relative to 
the $\gamma\gamma$ final state and where they stand vis-a-vis the experimental 
results. Such a study of relative strengths over the parameter space of our 
scenario would be indicative of how well the same is compatible with the 
experimental observations in the Higgs sector, in a global sense. The recent 
results from the ATLAS and the CMS collaborations on different decay modes of 
the lightest Higgs boson are presented in table \ref{signal_strengths}.
In fig. \ref{fig:comp-signal-strength}, we present the $\mu$-values reported 
by the ATLAS and the CMS collaborations for different final states in the 
so-called signature (ratio) space, in reference to $\mu_{\gamma \gamma}$.
\begin{table}[ht]
\centering
\begin{tabular}{|c|c|c|}
\hline
Channel & $\mu$ (CMS) & $\mu$ (ATLAS) \\ [0.5ex]
\hline
$h~\rightarrow\gamma\gamma$ & $1.14^{+~0.26}_
{-~0.23}$\cite{CMS-mugg-2} & $1.17^{+~0.27}_{-~0.27}$
\cite{ATLAS-mugg-1} \\
$h~\xrightarrow {ZZ^{*}} 4l$ & $0.93^{+~0.39}_
{-~0.32}$\cite{CMS-Z} & $1.44^{+~0.40}_{-0.33}$
\cite{ATLAS-mugg-1} \\
$h~\xrightarrow {WW^{*}} 2l 2\nu$ & $0.72^{+~0.20}
_{-~0.18}$ \cite{CMS-W}& $1.0^{+0.30}_{-0.30}$
\cite{ATLAS-W} \\
$h\rightarrow b\overline{b}$ & $1.0^{+~0.5}_{-~0.5}$
\cite{CMS-b} & $0.2^{+~0.70}_{-~0.60}$
\cite{ATLAS-b} \\
$h\rightarrow \tau\overline\tau$ & $0.78^{+~0.27}_{-~0.27}$
\cite{CMS-tau} & $1.4^{+~0.5}_{-~0.4}$\cite{ATLAS-tau} \\
\hline
\end{tabular}
\caption{\label{signal_strengths} Signal strengths ($\mu$) in different 
decay final states of the SM-like Higgs boson as reported by the CMS and the 
ATLAS collaborations (with the corresponding references). }
\end{table}

In each plot, the blue circle (green square)  
represents the experimentally reported central values for a given 
pair of observables from ATLAS and CMS collaborations, respectively. The solid grey lines show 
the range of $\mu$ values as observed by the CMS experiment while the dashed ones 
delineate the same as obtained by the ATLAS experiment. 
In order to generate fig. \ref{fig:comp-signal-strength} we vary $\tan\beta$ within
the range $10<\tan\beta<40$. We have also varied the mass of the top squark within
the range $350~ {\mathrm{GeV}} < m_{\widetilde t} < 1.5~ \rm{TeV}$ with 
$0.1 < f < 1$ and $0.1 < \lambda_T < 0.55$. All other parameters are kept fixed 
at the previously mentioned values in section~\ref{hgammagamma}. While scanning, care has been taken
to reject spectra with tachyonic scalar states and to conform with the lower 
bound on the lightest chargino mass of $104$ GeV as obtained from the LEP 
experiment. Also, the scan required $m_h$ to be within the range of 
$124.0-126.2$ GeV as reported by the LHC experiments. 
%
The spread in the upper two plots in fig. \ref{fig:comp-signal-strength}
are due to the variation of $f$ which affects $\mu_{bb}$ and $\mu_{\gamma\gamma}$
whereas $\mu_{WW}$ and $\mu_{ZZ}$ remain unaffected. The values of $\mu_{\gamma
\gamma}$ is very much consistent with the recent ATLAS and CMS findings.
Finally, in order to have an idea of the mass-spectra of the light 
neutralino and the chargino states, we provide a few benchmark points 
in table \ref{Tab:Spec1}, for the large `$f$' scenario.
\begin{table}[!htbp]
\centering
\begin{tabular}{|c|c|c|c|}
\hline
Parameters              & BP-$\it 1$           & BP-$\it 2$                 & BP-$\it 3$\\ [0.5ex]
\hline
$M_1^D$                 & 1500 GeV             & 1000 GeV             & 1200 GeV\\
$M_2^D$                 & 1500.1 GeV           & 1000.1 GeV           & 1200.1 GeV\\
$\mu_u$                 & 200 GeV              & 200 GeV              & 200 GeV\\
$m_{3/2}$               & 20 GeV               & 20 GeV               & 10 GeV\\
$\tan\beta$             & 25                   & 35                   & 40\\
$m_{\widetilde t}$          & 500 GeV              & 400 GeV              & 400 GeV\\
$f$                     & 0.8                  & 0.8                  & 0.8\\
$\lambda_T$             & 0.5                  & 0.52                 & 0.52\\
$v_S$                   & $10^{-4}$ GeV        & $10^{-4}$ GeV        & $10^{-4}$ GeV \\
$v_T$                   & $10^{-3}$ GeV        & $10^{-3}$ GeV        & $10^{-3}$ GeV\\
$B\mu_{L}$              & $-(400)^2$ (GeV)$^2$ & $-(400)^2$ (GeV)$^2$ & $-(400)^2$ (GeV)$^2$ \\
\hline
Observables              & BP-$\it 1$          & BP-$\it 2$                 & BP-$\it 3$\\ [0.5ex]
\hline
$m_h$                   & 124.98 GeV           & 125.45 GeV           & 125.73 GeV\\
$(m_{\nu})_{\rm{Tree}}$ & 0.04 eV              & 0.1 eV               & 0.08 eV\\
$m_{\widetilde\chi_7^0}$& 168 MeV              & 169 MeV              & 84 MeV\\
$m_{\widetilde\chi_6^0}$& 208.73 GeV          & 210.58 GeV          & 209.75 GeV\\
$m_{\widetilde\chi_5^0}$& 208.74 GeV          & 210.59 GeV          & 209.76 GeV\\
$m_{\widetilde\chi_4^0}$& 1504.17 GeV          & 1006.13 GeV          & 1205.29 GeV\\
$m_{\widetilde\chi_3^0}$& 1504.23 GeV          & 1006.19 GeV          & 1205.31 GeV\\
$m_{\widetilde\chi_2^0}$& 1.19$\times 10^5$ GeV& 1.11$\times 10^5$ GeV& $1.33\times 10^5$ GeV\\
$m_{\widetilde\chi_1^0}$& 1.19$\times 10^5$ GeV& 1.11$\times 10^5$ GeV& $1.33\times 10^5$ GeV\\
$m_{\widetilde\chi_3^+}$& 208.13 GeV           & 211.91 GeV          & 210.24 GeV\\
$m_{\widetilde\chi_2^+}$& 1500.11 GeV          & 1000.11 GeV          & 1200.1 GeV\\
$m_{\widetilde\chi_1^+}$& 1508.27 GeV          & 1012.15 GeV          & 1210.45 GeV\\
\hline
$\mu_{\gamma\gamma}$    & 1.07                 & 1.11                  & 1.11\\
\hline
\end{tabular}
\caption{\label{Tab:Spec1} Benchmark sets of input parameters in the
large Yukawa coupling ($f$) scenario and the
resulting mass-values for some relevant excitations. The Higgs signal
strength in the diphoton final state ($\mu_{\gamma \gamma}$) is also
indicated.
}
\end{table}

\subsection{The case of small Yukawa coupling, $f\sim \mathcal O(10^{-4})$}
In the limit when the Yukawa coupling is small ($f\sim 10^{-4}$), the 
next-to-lightest neutralino state becomes the sterile neutrino with negligible 
active-sterile mixing. The lightest neutralino state is again the active neutrino. 
The tree level Majorana mass of the active neutrino is given by 
eq.~(\ref{neutrino_majorana}) whereas the sterile neutrino mass and the mixing 
angle between the active and the sterile neutrino are given by 
eqs.~(\ref{sterile-mass}) and (\ref{sterile-mixing}). 
We have mentioned in the previous section that an X-ray line at around 3.5 keV was observed in the X-ray
spectra of the Andromeda galaxy and in the same from various other galaxy 
clusters including the Perseus cluster. The observed flux and the best fit 
energy peak are shown in \cite{Bulbul,Boyarsky-1}.
The origin of this line is disputed since atomic transitions in the thermal plasma
may also be responsible for this energy line. Nevertheless, a possible explanation 
can be provided by taking into account a 7 keV dark matter, in this case a sterile neutrino 
\cite{Bulbul,Boyarsky-1}. As discussed earlier, the observed flux 
and the peak of the energy can be 
translated to an active-sterile mixing in the range $2.2\times 10^{-11}
< \sin^2 2\theta_{14} < 2\times 10^{-10}$. To satisfy such small active sterile
mixing, the tree level neutrino mass turns out to be very small ($\mathcal O(10^{-5})$
eV). Therefore, in order to explain the neutrino mass and mixing, one needs to
invoke radiative corrections. For a detailed discussion, we refer the reader
to~\cite{Chakraborty}.
It is also important to study the 
signal strength of $h\rightarrow \gamma\gamma$ in the light of this 7 keV 
sterile neutrino with appropriate active-sterile mixing.
\begin{figure}[htb]
\begin{center}
\includegraphics[height=3.3in,width=3.8in]{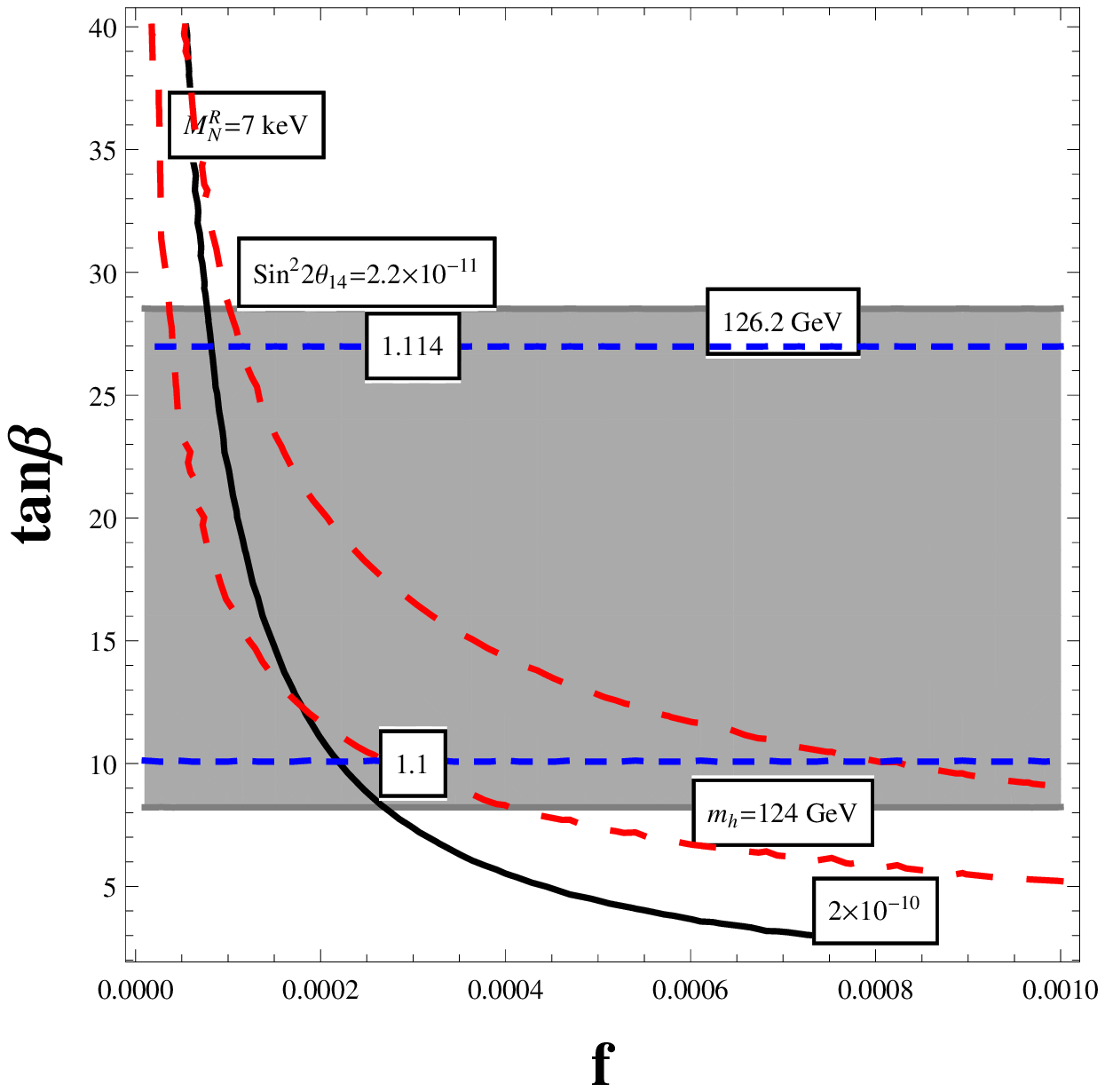}
\caption{\label{mugg_dm}
Contours of fixed values of $m_h$, $\mu_{\gamma\gamma}$, $M_N^R$ and 
$\sin^2 2\theta_{14}$ in the $f-\tan\beta$ parameter space. The respective 
values of the contour lines are as shown in the figure. The shaded region in 
grey corresponds to the experimentally allowed band of the lightest Higgs boson 
mass. Other parameters are fixed at values mentioned in the text.  
}
\end{center}
\end{figure}

\begin{figure}[htb]
\begin{center}
\includegraphics[height=3.3in,width=3.8in]{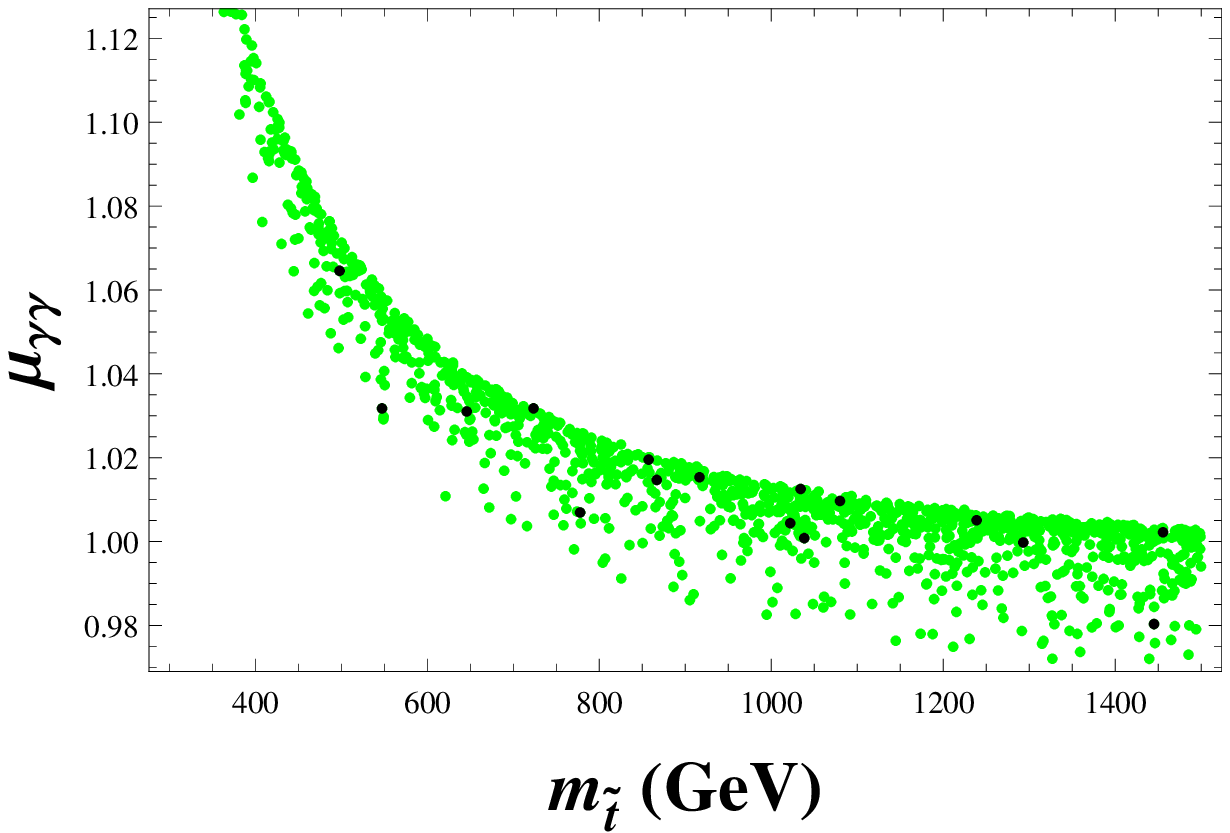}
\caption{\label{scatter_dm}
Scatter plot showing possible range of variation of
$\mu_{\gamma \gamma}$ with varying $m_{\widetilde{t}}$.
The blue points are consistent with $7.01~\rm{keV} < M_N^R < 7.11~\rm{keV}$. 
All points satisfy 124.0 GeV $< m_h <$ 126.2 GeV. 
}
\end{center}
\end{figure}

In fig. \ref{mugg_dm} we present the contours of $m_h$, $\mu_{\gamma\gamma}$, 
$M_N^R$ and $\sin^2 2\theta_{14}$ in the $f$--$\tan\beta$ plane. The contour of 
the sterile neutrino mass of 7 keV is shown with the thick black line. The red
dashed lines represent the contours of active-sterile mixing fixed at $2.2\times
10^{-11}$ and $2\times 10^{-10}$. We have fixed $M_1^D$ at 1 TeV, maintaining
a degeneracy $\epsilon = (M_2^D - M_1^D)=10^{-4}$ GeV. $\mu_u$ is fixed at 
500 GeV. The other fixed parameters are $m_{3/2} = 10~\rm{GeV}$, $m_{\widetilde t}=
400~\rm{GeV}$, $\lambda_T = 0.57$, $v_S = -0.01~\rm{GeV}$, $v_T = 0.01~
\rm{GeV}$ and $B\mu_L=-(400)^2$ (GeV)$^2$. The not so heavy top squark, as 
justified in section \ref{sec:signal-strength}, enhances $\mu_{\gamma\gamma}$ 
considerably and we show the contours of $\mu_{\gamma\gamma}$ at 1.1 and 1.114  
respectively with blue dashed lines. Finally, the grey shaded region is 
the parameter space consistent with the observed Higgs boson mass 
$124.0~{\mathrm {GeV}} < m_h < 126.2~{\mathrm{GeV}}$. Figure 
\ref{mugg_dm} clearly shows that for this choice of parameters 
$\mu_{\gamma\gamma}\gsim 1.1$ is completely consistent with a 7 keV sterile 
neutrino dark matter and the experimentally allowed range of Higgs boson 
mass. 
We have seen that charginos do not provide much enhancement to 
$\mu_{\gamma\gamma}$ due to its very suppressed couplings under the 
present set-up. Furthermore, avoiding possible appearance 
of tachyonic scalar states restricts the $vev$ of the singlet from becoming 
large. Therefore, expecting an enhancement in $\mu_{\gamma \gamma}$ 
via suppression of the $hb{\overline b}$ coupling because of the singlet admixture seems unrealistic. 
Thus, the only enhancement in $\mu_{\gamma \gamma}$ can come from light top 
squarks. In addition, large radiative corrections from $\lambda_S$ and 
$\lambda_T$ reduces the necessity of having heavy top squarks. 
In the scatter plot of fig. 
\ref{scatter_dm} we show the possible range of variation of 
$\mu_{\gamma\gamma}$ with varying $m_{\widetilde t}$.  
To generate this plot we have chosen relevant parameters over the following 
ranges: $1~{\mathrm{GeV}} < m_{3/2} < 20~{\mathrm{GeV}}$, $5 < \tan\beta < 40$, 
$300~{\mathrm{GeV}} < m_{\widetilde t} < 1.5~{\mathrm{TeV}}$, $10^{-5} < f < 3\times 
10^{-4}$, $0.1 < \lambda_T < 1$ and $-0.01~{\mathrm{GeV}} < v_S < 
-1~{\mathrm{GeV}}$. Other parameters are retained at their previously mentioned 
values (used to obtain fig. \ref{mugg_dm}), maintaining the degeneracy between 
the Dirac gaugino masses as already 
mentioned. Again, all these points are consistent with $124.0~{\mathrm{GeV}} 
< m_h < 126.2~{\mathrm{GeV}}$ and free from any tachyonic scalar states. 
The effects of the light top squarks results in some enhancement in 
$\mu_{\gamma \gamma}$. The blue points are consistent
with a keV sterile neutrino with mass ranging between $7.01~\rm{keV} < M_N^R 
< 7.11~\rm{keV}$ and is known to be a fit warm dark matter candidate having the 
right relic density. Finally, it is again very relevant to check the relative signal strengths for 
different decay modes of the lightest Higgs boson in such a scenario with small 
`$f$'; similar to what we have done in section \ref{sec:relative-strength-large-f} for 
the large `$f$' scenario.  
\begin{figure}%
    \centering
    {{\includegraphics[width=7cm]
    {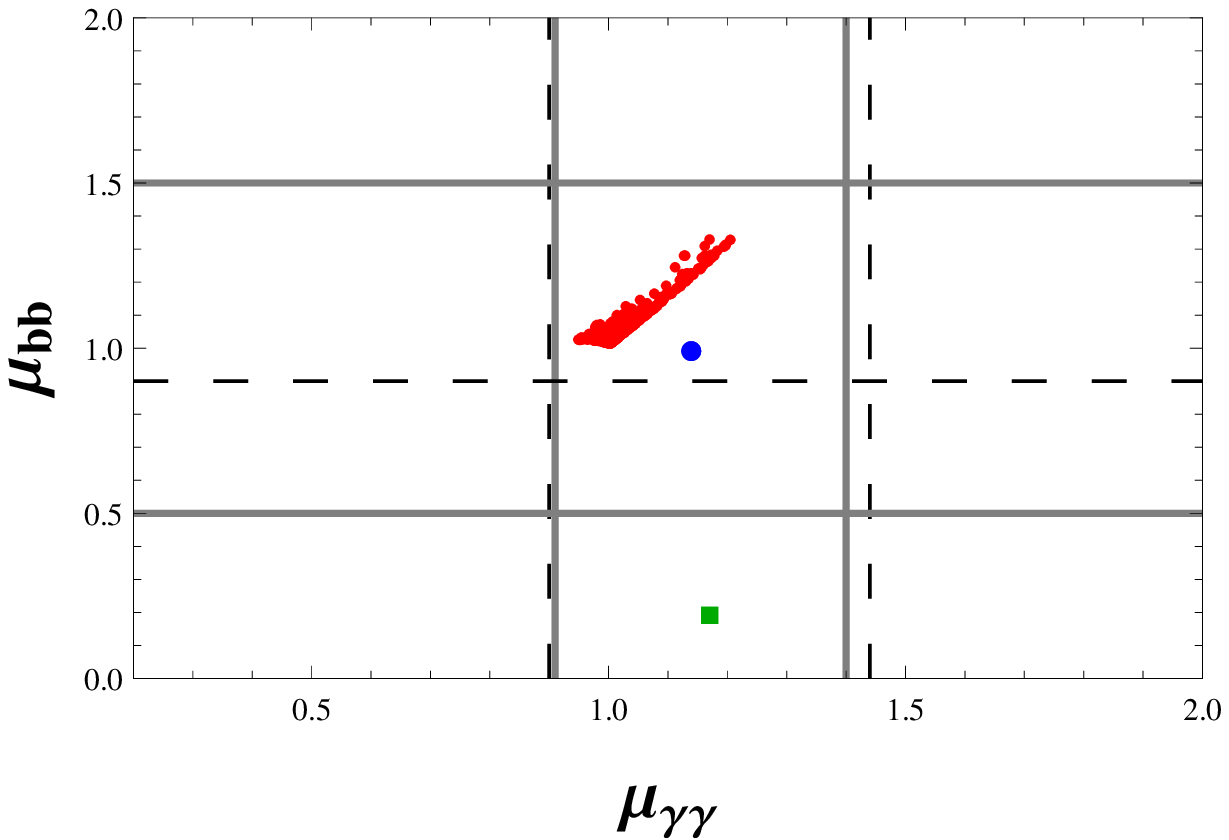} }}%
    \qquad
    {{\includegraphics[width=7cm]
    {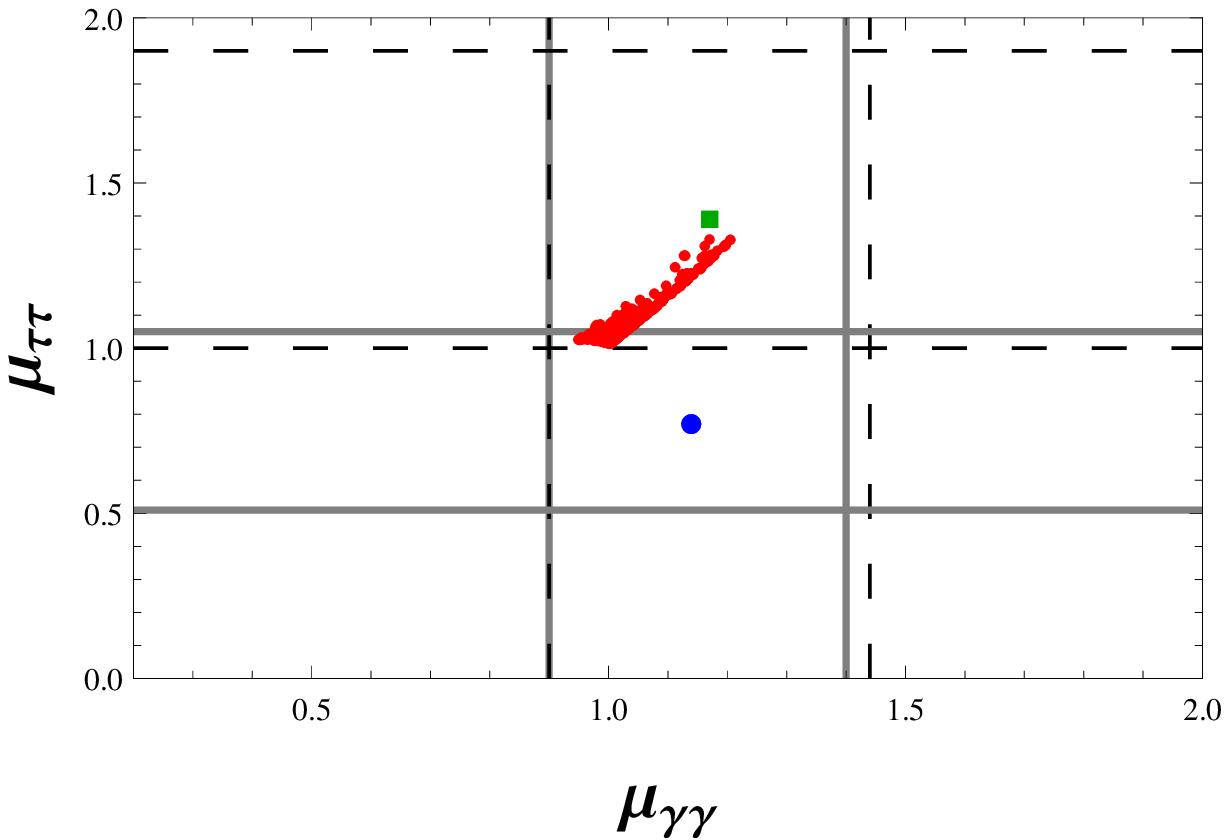} }}%
    \qquad
    {{\includegraphics[width=7cm]
    {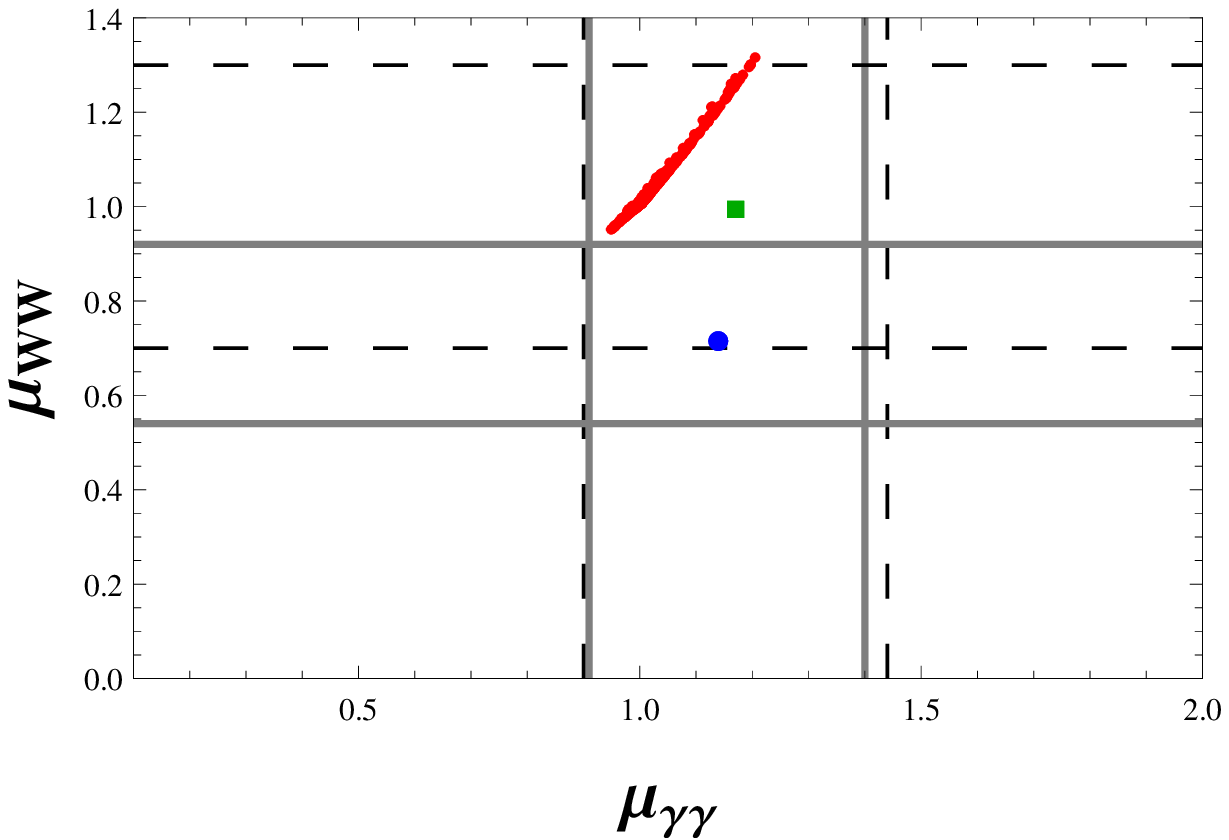} }}%
    \qquad
    {{\includegraphics[width=7cm]
    {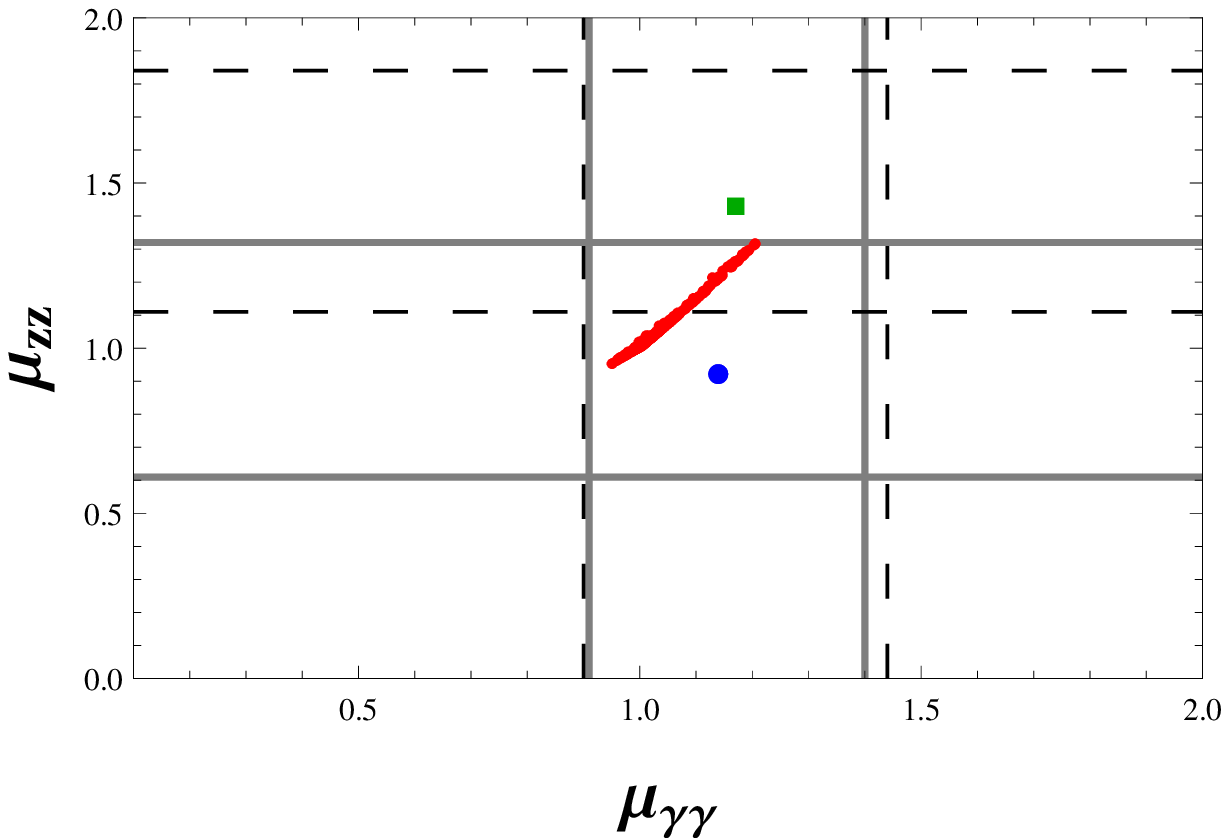} }}%
    \caption{Same as in figure \ref{fig:comp-signal-strength} except for a small 
    input value of $f$.  
    }
    \label{comp-signal-strength-dm}%
\end{figure}
Figure \ref{comp-signal-strength-dm} shows scattered points consistent with the 
CMS or/and the ATLAS results at 1$\sigma$ level. However, note that the scatter 
plot in the $\mu_{\gamma \gamma}$--$\mu_{WW}$ plane is consistent only with the 
results from the ATLAS experiments at the 1$\sigma$ level whereas the the scatter
plot in the $\mu_{\gamma \gamma}$--$\mu_{bb}$ plane is consistent only with the
results from the CMS experiments at the 1$\sigma$ level. In the near future, 
a more precise measurement together with an improved analysis is likely to 
become more decisive on this issue. 
%
Finally, for the sake of completeness, in table~\ref{Tab:Spec2} we 
provide three more benchmark sets comprising of the input parameters 
of the small Yukawa coupling scenario (with $(f\sim 10^{-4})$), the 
corresponding mass-values of the relevant excitations and the Higgs signal 
strengths in the diphoton final state ($\mu_{\gamma \gamma}$).
\begin{table}[!htbp]
\centering
\begin{tabular}{|c|c|c|c|}
\hline
Parameters              & BP-$\it 4$            & BP-$\it 5$                 & BP-$\it 6$\\ [0.5ex]
\hline
$M_1^D$                 & 1000 GeV              & 900 GeV              & 1200 GeV\\
$\mu_u$                 & 300 GeV               & 600 GeV              & 600 GeV\\
$m_{3/2}$               & 4 GeV                 & 10 GeV               & 15 GeV\\
$\tan\beta$             & 35                    & 25                   & 15\\
$m_{\widetilde t}$          & 500 GeV               & 500 GeV              & 500 GeV\\
$f$                     & 9.9$\times 10^{-5}$   & 8.9$\times 10^{-5}$  & 1.21$\times10^{-4}$\\
$\lambda_T$             & 0.55                  & 0.55                 & 0.55\\
$v_S$                   & -$10^{-2}$ GeV        & -$10^{-2}$ GeV       & -$10^{-2}$ GeV \\
$v_T$                   & $10^{-2}$ GeV         & $10^{-2}$ GeV        & $10^{-2}$ GeV\\
\hline
Observables              & BP-$\it 4$           & BP-$\it 5$                 & BP-$\it 6$\\ [0.5ex]
\hline
$m_h$                   & 125 GeV               & 124.257 GeV          & 124.448 GeV\\
$m_N^R$                 & 7.03 keV              & 7.09 keV             & 7.03 keV\\
$m_{\widetilde\chi_6^0}$& 292.375 GeV           & 571.91 GeV          & 587.24 GeV\\
$m_{\widetilde\chi_5^0}$& 292.376 GeV           & 571.92 GeV           & 587.25 GeV\\
$m_{\widetilde\chi_4^0}$& 1004.06 GeV           & 904.16 GeV          & 1203.24 GeV\\
$m_{\widetilde\chi_3^0}$& 1004.07 GeV           & 904.19 GeV          & 1203.28 GeV\\
$m_{\widetilde\chi_2^0}$& 1022.03 GeV           & 939.91 GeV          & 1222.84 GeV\\
$m_{\widetilde\chi_1^0}$& 1022.72 GeV           & 939.83 GeV          & 1222.72 GeV\\
$m_{\widetilde\chi_3^+}$& 311.56 GeV           & 609.77 GeV          & 608.27 GeV\\
$m_{\widetilde\chi_2^+}$& 1000.01 GeV           & 900.01 GeV          & 1200.02 GeV\\
$m_{\widetilde\chi_1^+}$& 1011.93 GeV           & 910.62 GeV          & 1208.7 GeV\\
$\sin^2 2\theta_{14}$   & $1.56\times 10^{-10}$ & $4.7\times 10^{-11}$ & $2.8\times 10^{-11}$ \\
\hline
$\mu_{\gamma\gamma}$    & 1.07                  & 1.06                  & 1.06\\
\hline
\end{tabular}
\caption{\label{Tab:Spec2} Same as in table \ref{Tab:Spec1} but for
small Yukawa coupling with $f \sim\mathcal{O}(10^{-4})$. 
In all three cases we have chosen $\epsilon = 10^{-4}$ GeV. 
Neutrino mass at the tree level is very small ($\mathcal {O}(10^{-5})$ eV)
and not shown in the table (See text for more details).
}
\end{table}
\section{Concluding remarks}
\label{chap7-concl}
In this paper we study the $h\rightarrow\gamma\gamma$ channel in the $U(1)_R$
lepton number model with a right handed neutrino. We show that the recent results
from ATLAS and CMS on $\mu_{\gamma\gamma}$ is very much consistent with our
outcomes for both the cases, i.e., $f\sim\mathcal O(1)$ and $f\sim\mathcal O(10^{-4})$.
We also show for large neutrino Yukawa coupling, $f$ the light bino-like neutralino 
state is not yet constrained from the invisible branching fraction of the Higgs boson.

So far we have seen that the model under consideration have already
demonstrated its ability to attract constraints from recent experiments
in diverse areas ranging from the neutrino to astro-particle physics and
finally from the LHC experiments pertaining to the Higgs sector and other BSM
searches. It will be really interesting to see if the model can provide any novel
signatures as far as the collider experiments are concerned.  
\appendix
\section{The Higgs-chargino-chargino coupling}
\label{app:hchch}
In this appendix we work out the Higgs-chargino-chargino coupling in the
scenario under discussion and present the analytical expression for the width
of the lightest Higgs boson decaying into a pair of charginos. The relevant 
Lagrangian in the two-component notation containing the Higgs-chargino-chargino 
interaction is given by
\bea
\mathcal L_{h\widetilde\chi^{+} \widetilde\chi^{-}}&=& g\left(v_a + \frac{S_{i2}}
{\sqrt 2} h_i\right)\widetilde w^+ e_L^- 
+ \sqrt 2 \lambda_T \widetilde T_u^+ \left(v_u + 
\frac{S_{i1}}{\sqrt 2} h_i\right) \widetilde R_d^- \nonumber \\ 
&+& g \left(v_u + \frac{S_{i1}}
{\sqrt 2}h_i\right)\widetilde H_u^+ \widetilde w^- 
-\lambda_S \left(v_S + \frac{S_{i3}}
{\sqrt 2} h_i\right)\widetilde H_u^+ \widetilde R_d^- \nonumber \\
&+&\lambda_T \left(v_T + \frac{S_{i4}}{\sqrt 2} h_i\right)\widetilde H_u^+ 
\widetilde R_d^- + g\left(v_T + \frac{S_{i4}}{\sqrt 2} h_i\right) \tilde T_u^+
\tilde w^- \nonumber \\
&-& g \left(v_T + \frac{S_{i4}}{\sqrt 2} h_i\right) \tilde w^+ \tilde T_d^- +h.c.,
\label{chargino-lagrangian}
\eea 
where the matrix $S$ connects the mass and gauge eigenstates of the CP even 
scalar mass squared matrix, written in the basis ($h_R$, $\tilde \nu_R$, $S_R$, 
$T_R$).
To be more precise the physical CP-even scalar states are related to the gauge 
eigenstates in the following manner:
\bea
\begin{pmatrix}
h_1 \\
h_2 \\
h_3 \\
h_4
\end{pmatrix} =
\left(\begin{array}{cccc}
S_{11} & S_{12} & S_{13} & S_{14} \\
S_{21} & S_{22} & S_{23} & S_{24} \\
S_{31} & S_{32} & S_{33} & S_{34} \\
S_{41} & S_{42} & S_{43} & S_{44} 
\end{array} \right)
\begin{pmatrix}
h_R \\
\tilde \nu_R \\
S_R \\
T_R
\end{pmatrix}.
\eea
In our notation the lightest physical state ($h_4$) of the CP even scalar mass 
matrix corresponds to the physical Higgs boson, $h$. Moreover, the charginos 
$\tilde \chi_i^\pm$ are four component Dirac fermions which arise due to the 
mixing between the charged gauginos and higgsinos as well as the charged lepton 
of first generation. In order to evaluate find out the Higgs-chargino-chargino 
coupling and to evaluate the Higgs boson partial decay width to a pair of 
charginos, it is pertinent to write down the interaction Lagrangian in the 
four-component notation. We now define the 4-component spinors as
\bea
\widetilde W = \begin{pmatrix}
\widetilde w^+ \\
{\bar {\widetilde w}}^-
\end{pmatrix},
\hskip 0.2cm
\widetilde H = \begin{pmatrix}
\widetilde H_u^+ \\
{\bar {\widetilde R}}_d^-
\end{pmatrix},
\hskip 0.2cm
\widetilde T = \begin{pmatrix}
\widetilde T_u^+ \\
{\bar {\widetilde T}}_d^-
\end{pmatrix},
\hskip 0.2cm
L_e^{(4)} = \begin{pmatrix}
e_R^c \\
\bar e_L^-
\end{pmatrix}.
\eea
Using the transformation relations, 
\bea
\widetilde w^+ e_L^- &=& \bar L_e^{(4)} P_L \widetilde W \nonumber \\
\widetilde T_u^+ \widetilde R_d^- &=& \overline{\widetilde H} P_L 
\widetilde T \nonumber \\
\widetilde H_u^+ \widetilde w^- &=& \overline{\widetilde W} P_L 
\widetilde H \nonumber \\
\widetilde H_u^+ \widetilde R_d^- &=&\overline{\widetilde H}P_L \widetilde H,
\eea
the Lagrangian in eq.~({\ref{chargino-lagrangian}}) can be expressed in the 
four component notation as 
\bea
\mathcal L_{h \widetilde\chi^+ \widetilde\chi^-}^{(4)} &=& g \frac{S_{42}}
{\sqrt 2} 
h \overline L_e^{(4)} P_L \widetilde W + \sqrt 2\lambda_T \frac{S_{41}}
{\sqrt 2} h 
\overline{\widetilde H}P_L \widetilde T  
+ g \frac{S_{41}}{\sqrt 2} h \overline{\widetilde W} P_L \widetilde H 
-\lambda_S \frac{S_{43}}{\sqrt 2} h 
\overline{\widetilde H}P_L \widetilde H \nonumber \\ 
&+& \lambda_T \frac{S_{44}}{\sqrt 2} h \overline{\widetilde H} P_L 
\widetilde H  + g \frac{S_{44}}{\sqrt 2} h \overline{\widetilde W} P_L \widetilde T
- g \frac{S_{44}}{\sqrt 2}\overline{\widetilde T} P_L \widetilde W + h.c. 
\label{chargino-gauge}
\eea
The chargino masses can have any sign. By demanding that the 
four component Lagrangian contains only positive masses for the charginos,
we define the chargino states in the following manner \cite{Gun,Gun-1}
\bea
\widetilde \chi^+_i &=& (\epsilon_i P_L + P_R) \begin{pmatrix}
\chi_i^+ \\
\bar{\chi}_i^-
\end{pmatrix}, ~~~~~~i = 1,...,4
\eea
where $\epsilon_i$ carries the sign of the chargino masses, which can be $\pm 1$.
When $\epsilon = -1$, $P_R - P_L = \gamma_5$, which essentially implies a 
$\gamma_5$ rotation to the four component spinors to absorb the sign. Hence, the 
transformation relations involving only $P_L$ changes, which modifies the 
Feynman rules. The two-component mass eigenstates ($\chi_i^\pm$) of the charginos 
are related to the gauge eigenstates in a manner shown in 
eq.~(\ref{chargino-mass-two}).
 
Using the following set of relations
\bea
P_L \widetilde W &=& P_L V_{i1}^* \epsilon_i \widetilde \chi_i \nonumber \\
P_L \widetilde T &=& P_L V_{i2}^* \epsilon_i \widetilde \chi_i \nonumber \\
P_L \widetilde H &=& P_L V_{i3}^* \epsilon_i \widetilde \chi_i \nonumber \\
P_R \widetilde W &=& P_R U_{i1} \widetilde \chi_i \nonumber \\
P_R \widetilde H &=& P_R U_{i3} \widetilde \chi_i \nonumber \\
P_R \widetilde T &=& P_R U_{i2} \widetilde \chi_i \nonumber \\
P_R L_e^{(4)} &=& P_R U_{i4} \widetilde\chi_i ,
\eea
we rewrite eq.~(\ref{chargino-gauge}) in the mass eigenstate basis as
\bea
\mathcal L_{h\tilde\chi_i^+ \tilde\chi_j^-}^{(4) m} &=& g h \overline{\tilde 
\chi}_i 
\left(\zeta_{ij}^{*} P_L + \zeta_{ji} P_R\right) \tilde \chi_j,
\label{lag-four}
\eea
\noindent where
\bea
\zeta_{ij} &=& \bigg[\frac{S_{42}}{\sqrt 2} U_{i4} V_{j1} + \sqrt 2 
\frac{\lambda_T}{g} 
\frac{S_{41}}{\sqrt 2} U_{i3} V_{j2} + \frac{S_{41}}{\sqrt 2} U_{i1} V_{j3} 
- \frac{\lambda_S}{g} \frac{S_{43}}{\sqrt 2} U_{i3} V_{j3} \nonumber \\
&+& \frac{\lambda_T}{g} \frac{S_{44}}{\sqrt 2} U_{i3} V_{j3}+\frac{S_{44}}{\sqrt 2} 
U_{i1}V_{j2} - \frac{S_{44}}{\sqrt 2} U_{i2} V_{j1}\bigg]\epsilon_i. \nonumber \\
\eea
\begin{figure}[htb]
\begin{center}
\includegraphics[height=1.8in,width=2.8in]
{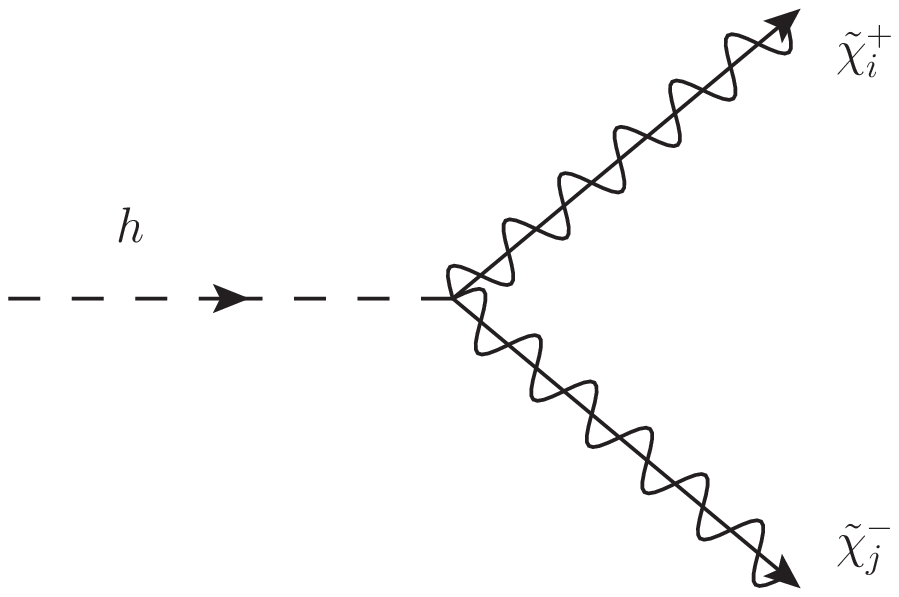}
\caption{\label{higgs-chargino-chargino}
The Higgs-chargino-chargino vertex.}
\end{center}
\end{figure}
The coupling is obtained from Eq.~(\ref{lag-four}) as 
\bea
\frac{g}{2}\left[\zeta_{ij}^* (1-\gamma_5) + \zeta_{ji}(1+\gamma_5) \right].
\eea
It is now straightforward to compute the lightest Higgs boson decay width to a 
pair of charginos, which we find as
\bea
\Gamma_{h\rightarrow \widetilde\chi_i^{+}\widetilde\chi_j^{-}} &=& \frac{g^2}
{16\pi m_h^3} 
\Big[\big\{m_h^2 - (m^2_{\widetilde\chi_i^+} + m^2_{\widetilde\chi_j^-})
\big\}^2  
- 4 m^2_{\widetilde\chi_i^+} m^2_{\widetilde\chi_j^-}\Big]^{1/2}  \nonumber \\
&&\Big[(\zeta_{ij}^2 + \zeta_{ji}^2) 
(m_h^2 - m^2_{\widetilde\chi_i^+} -  m^2_{\widetilde\chi_j^-}) 
- 4 \zeta_{ij} \zeta_{ji} m_{\widetilde\chi_i^+} m_{\widetilde\chi_j^-}\Big].
\eea
Finally, if we assume the singlet and the triplet $vev$'s to be very small,
this would imply that the singlet and triplet mixing in the light 
CP-even Higgs boson states become negligible. 
Under such an assumption, the CP even states can be written as
\bea
\tilde\nu_R &\simeq& v_a + \frac{1}{\sqrt 2} \left(H \cos\alpha - 
h\sin\alpha\right) \nonumber \\
h_R &\simeq& v_u + \frac{1}{\sqrt 2} \left(H \sin\alpha +
h\cos\alpha\right),
\eea
where we have chosen $S_{41}=\cos\alpha$,
$S_{42}=-\sin\alpha$, and $S_{43}\sim S_{44}\sim 0$. With this
simplification we can write
\bea
\zeta_{ij} &=& \bigg[-\frac{\sin\alpha}{\sqrt 2}U_{i4}V_{j1}
+\frac{\cos\alpha}{\sqrt 2}\bigg(\frac{\sqrt 2\lambda_T}{g} U_{i3} V_{j2}
+ U_{i1} V_{j3}\bigg)\bigg]\epsilon_i \nonumber \\
&=& \xi_{ij}\sin\alpha - \eta_{ij} \cos\alpha,
\eea
where
\bea 
\xi_{ij} &=& -\frac{U_{i4}V_{j1}}{\sqrt 2} \epsilon_i \nonumber \\
\eta_{ij} &=& \frac{1}{\sqrt 2} \bigg(\frac{\sqrt 2\lambda_T}{g} U_{i3} V_{j2} 
+ U_{i1} V_{j3} \bigg) \epsilon_i.
\eea
%
\section{The Higgs-neutralino-neutralino coupling}
\label{app:hntnt}
In a similar manner the interaction of the Higgs boson with neutralinos can be
constructed from the following (two-component) Lagrangian
\bea
\mathcal L_{h\widetilde\chi^0\widetilde\chi^0} &=& \frac{g^{\prime}}{\sqrt 2}
\left(
v_u + \frac{S_{i1}}{\sqrt 2} h_i\right) \widetilde b\widetilde H_u^0 
- \frac{g^{\prime}} {\sqrt 2}\left(v_a + \frac{S_{i2}}{\sqrt 2} h_i\right)
\widetilde b \nu_e 
+ \lambda_S \left(v_u + \frac{S_{i1}}{\sqrt 2} h_i\right)\widetilde S
\widetilde R_d^0 \nonumber \\
&-& \frac{g}{\sqrt 2} \left(v_u + \frac{S_{i1}}{\sqrt 2} h_i\right)\widetilde 
w \widetilde H_u^0 
+ \frac{g}{\sqrt 2}\left(v_a + \frac{S_{i2}}{\sqrt 2} h_i\right)\widetilde 
w \nu_e
+ \lambda_T \left(v_u + \frac{S_{i1}}{\sqrt 2} h_i\right)\widetilde T\widetilde 
R_d^0 \nonumber \\
&+& \left[\lambda_S \left(v_s + \frac{S_{i3}}{\sqrt 2} h_i\right)
+ \lambda_T \left(v_T + \frac{S_{i4}}{\sqrt 2} h_i\right)\right]
\widetilde R_d^0 \widetilde H_u^0 
- f \left(v_a + \frac{S_{i2}}{\sqrt 2}h_i\right)\widetilde H_u^0 N^c \nonumber \\
&-& f \left(v_u + \frac{S_{i1}}{\sqrt 2}h_i\right)N^c \nu_e + h.c. \nonumber \\
\eea
We stick to the notation for the lightest CP even physical scalar state being 
denoted by {$h_4$} and identified with the lightest Higgs boson $h$. We again 
define the 4-component spinors as \cite{Proy}
\bea
\tilde B &=& \begin{pmatrix}
\tilde b \\
\bar {\widetilde b}^{T}
\end{pmatrix},
\hskip 0.05cm
\tilde S = \begin{pmatrix}
\tilde S \\
\bar {\widetilde S}^{T}
\end{pmatrix},
\hskip 0.05cm
\tilde R_d = \begin{pmatrix}
\tilde R_d^0 \\
\bar {\widetilde R}_d^{0 T}
\end{pmatrix},
\hskip 0.05cm
\tilde H_u = \begin{pmatrix}
\tilde H_u^0 \\
\bar {\widetilde H}_u^{0 T}
\end{pmatrix},
\hskip 0.05cm
\nonumber \\
\tilde T &=& \begin{pmatrix}
\tilde T \\
\bar {\widetilde T}^{T}
\end{pmatrix},
\hskip 0.05cm
\tilde W = \begin{pmatrix}
\tilde W \\
\bar {\widetilde W}^{T}
\end{pmatrix},
\hskip 0.05cm
\nu_e = \begin{pmatrix}
\nu_e \\
\bar {\nu}_e^{T}
\end{pmatrix},
\hskip 0.05cm
N^c = \begin{pmatrix}
N^c \\
\bar {N}^{c T}
\end{pmatrix}.
\nonumber \\
\eea
In terms of these spinors the 4-component Lagrangian takes the following form
\bea
\mathcal L_{h\widetilde\chi^0 \widetilde\chi^0}^{(4)} &=& \frac{g^{\prime}}
{\sqrt 2}\frac{S_{41}}{\sqrt 2} h\bar{\widetilde B}P_L \widetilde H_u - 
\frac{g^{\prime}}{\sqrt 2}\frac{S_{42}}{\sqrt 2} h \bar{\widetilde B}P_L \nu_e 
+ \lambda_S \frac{S_{41}}{\sqrt 2} h \bar{\widetilde S}P_L \widetilde R_d 
- \frac{g}{\sqrt 2} \frac{S_{41}}{\sqrt 2}h \bar{\widetilde W} P_L 
\tilde H_u \nonumber \\ 
&+& \frac{g}{\sqrt 2} \frac{S_{42}}{\sqrt 2} h\bar {\widetilde W} P_L \nu_e
+ \lambda_T \frac{S_{41}}{\sqrt 2} h \bar{\widetilde T} P_L \widetilde R_d 
+ \lambda_S \frac{S_{43}}{\sqrt 2}h \bar{\widetilde R}_d P_L \widetilde H_u 
+ \lambda_T \frac{S_{44}} {\sqrt 2} h \bar{\widetilde R}_d P_L 
\widetilde H_u \nonumber \\
&-& f \frac{S_{42}}{\sqrt 2}h \bar{\widetilde H_u} P_L N^c
- f \frac{S_{41}}{\sqrt 2}h \bar{N}^c P_L \nu_e + h.c. \nonumber \\
\label{int-neutralino}
\eea
Eq.~(\ref{int-neutralino}) represents the interactions in the gauge eigenstate 
basis. Neutralinos are physical Majorana spinors, arising due to the mixing of 
the neutral gauginos, higgsinos as well as the active (first generation) and 
sterile neutrino states. The four component neutralino state is defined as
\bea
\widetilde \chi_i^0 &=& (\epsilon_i P_L + P_R)\begin{pmatrix}
\chi_i^0 \\
\bar{\chi}_i^0
\end{pmatrix}, ~~~~~~i = 1,...,8
\eea
where $\chi_i^0$ are two component neutralino mass eigenstates and they are
related to the gauge eigenstates as
\bea
\chi_i^0 &=& N_{ij} \psi_j^0, ~~~~~~i,j = 1,...,8
\eea 
where $\psi^0 = \big(\widetilde b, \widetilde S, \widetilde W, \widetilde T, 
\widetilde R_d, \widetilde H_u, N^c, \nu_e\big)^T$.
As presented in Appendix~A, in a similar fashion we use the following 
transformation relations to write down the interaction Lagrangian given in 
Eq.~(\ref{int-neutralino}) in the mass eigenstate basis 
\bea
P_L\widetilde B=N_{i1}^* P_L\epi\widetilde\chi_i^0 &,& P_R \widetilde B=N_{i1}
P_R\widetilde\chi_i^0\nonumber \\
P_L\widetilde S=N_{i2}^* P_L\epi\widetilde\chi_i^0 &,& P_R \widetilde S=N_{i2}
P_R\widetilde\chi_i^0\nonumber \\
P_L\widetilde W=N_{i3}^* P_L\epi\widetilde\chi_i^0 &,& P_R \widetilde W=N_{i3}
P_R\widetilde\chi_i^0\nonumber \\
P_L\widetilde T=N_{i4}^* P_L\epi\widetilde\chi_i^0 &,& P_R \widetilde T=N_{i4}
P_R\widetilde\chi_i^0\nonumber \\
P_L\widetilde R_d=N_{i5}^* P_L\epi\widetilde\chi_i^0 &,& P_R \widetilde R_d=N_{i5}
P_R\widetilde\chi_i^0\nonumber \\
P_L\widetilde H_u=N_{i6}^* P_L\epi\widetilde\chi_i^0 &,& P_R \widetilde H_u=N_{i6}
P_R\widetilde\chi_i^0\nonumber \\
P_L N^c=N_{i7}^* P_L\epi\widetilde\chi_i^0 &,& P_R N^c=N_{i7}
P_R\widetilde\chi_i^0\nonumber \\
P_L \nu_e=N_{i8}^* P_L\epi\widetilde\chi_i^0 &,& P_R  
\nu_e=N_{i8}P_R\widetilde\chi_i^0.
\nonumber \\
\eea
It is now straightforward to write down the Higgs-neutralino-neutralino 
interaction in the 4-component notation as
\bea
\mathcal L_{h\tilde\chi^0\tilde\chi^0}^{(4) m} &=& g \bar{\tilde\chi}_i^0 h 
\left(
\zeta_{ij}^{\prime *} P_L + \zeta_{ji}^{\prime} P_R\right)\tilde \chi_j^0,
\eea
where 
\bea
\zeta^{\prime}_{ij} &=& S_{41}\Big[\frac{g^{\prime}}{g} 
\frac{N_{i1}N_{j6}}{2}
+ \frac{\lambda_S}{g}\frac{N_{i2} N_{j5}}{\sqrt 2} - 
\frac{N_{i3}N_{j6}}{2} 
+ \frac{\lambda_T}{g} \frac{N_{i4} N_{j5}}{\sqrt 2}
- \frac{f}{g} \frac{N_{i7} N_{j8}}{\sqrt 2}\Big]\epsilon_i \nonumber \\
&+& S_{42} \Big[\frac{N_{i3} N_{j8}}{2} - \frac{g^{\prime}}{g} 
\frac{N_{i1}
N_{j8}}{2} - \frac{f}{g}\frac{N_{i6}N_{j7}}{\sqrt 2}\Big]\epsilon_i 
+ S_{43}\Big[\frac{\lambda_S}{g}\frac{N_{i5}N_{j6}}{\sqrt 2}\Big]
\epsilon_i \nonumber \\
&+& S_{44}\Big[\frac{\lambda_T}{g}\frac{N_{i5}N_{j6}}{\sqrt 2}\Big]
\epsilon_i+(i\leftrightarrow j). 
\eea

\begin{figure}[htb]
\begin{center}
\includegraphics[height=1.8in,width=2.8in]
{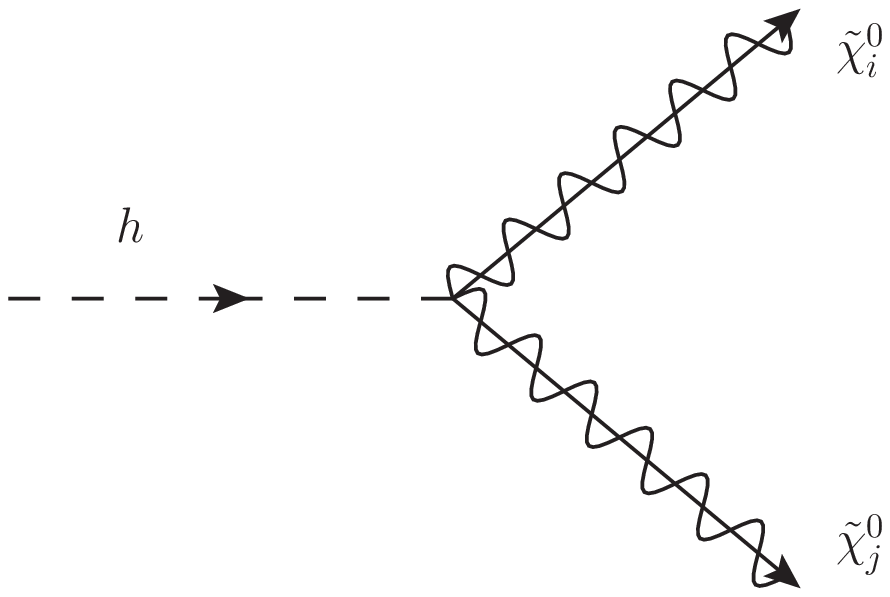}
\caption{\label{higgs-neutralino-neutralino}
The Higgs-neutralino-neutralino vertex.}
\end{center}
\end{figure}
Finally, the partial decay width $\Gamma(h\rightarrow \widetilde\chi_i^0 
\widetilde\chi_j^0)$ is given as
\bea
\Gamma_{h\rightarrow \widetilde\chi_i^0\widetilde\chi_j^0} &=& \frac{g^2}
{16\pi m_h^3 (1+\delta_{ij})}\Big[\{m_h^2 - (m^2_{\widetilde\chi_i^0} + m^2_{\widetilde
\chi_j^0})\}^2 - 4 m^2_{\widetilde\chi_i^0} m^2_{\widetilde\chi_j^0}\Big]^{1/2} 
\times \nonumber \\
&& \Big[\left(\zeta_{ij}^{\prime 2} + \zeta_{ji}^{\prime 2}\right) 
\left(m_h^2 - m^2_{\widetilde\chi_i^0} -m^2_{\widetilde\chi_j^0}\right) 
- 4\zeta_{ij} ^{\prime} \zeta_{ji}^{\prime} m_{\widetilde\chi_i^0} m_{\widetilde
\chi_j^0}\Big].
\eea
%
 
Again in the limit where the singlet and triplet $vev$'s are very small, we can 
safely ignore the contributions from $S_{43}$ and $S_{44}$. Furthermore,
replacing $S_{41}$ by $\cos\alpha$ and $S_{42}$ by -$\sin\alpha$, we can write
\bea
\zeta_{ij}^{\prime} &=& \eta_{ij}^{\prime} \cos\alpha  + \xi_{ij}^{\prime}
\sin\alpha,
\eea 
where,
\bea
\eta_{ij}^{\prime} &=& \bigg[\frac{g^{\prime}}{g} \frac{N_{i1}N_{j6}}{2}
+\frac{\lambda_S}{g}\frac{N_{i2}N_{j5}}{\sqrt 2} - \frac{N_{i3}N_{j6}}{2} 
+ \frac{\lambda_T}{g}\frac{N_{i4}N_{j5}}{\sqrt 2} - \frac{f}{g}
\frac{N_{i7}N_{j8}}{\sqrt 2}\bigg]\epsilon_i+(i\leftrightarrow j), \nonumber \\
\xi_{ij}^{\prime} &=& \bigg[\frac{g^{\prime}}{g} \frac{N_{i1}N_{j8}}{2} 
+ \frac{f}{g} \frac{N_{i6}N_{j7}}{\sqrt 2}-\frac{N_{i3}N_{j8}}{2}\bigg]
\epsilon_i+(i\leftrightarrow j).
\eea
\acknowledgments
SC would like to thank the Council of Scientific and Industrial Research, 
Government of India for the financial support received as a Senior Research 
Fellow. AD acknowledges the hospitality of the Department of Theoretical 
Physics, IACS during the course of this work. SR would like to thank the
hospitality of the University of Helsinki and Helsinki Institute of Physics
during the final stages of this work. It is also a pleasure to thank Dilip Kumar Ghosh,
Katri Huitu and Oleg Lebedev for helpful discussions. 

\end{document}